\documentclass[english,3p]{elsarticle}
\usepackage[T1]{fontenc}
\usepackage[utf8]{inputenc}
\usepackage{array}
\usepackage{verbatim}
\usepackage{float}
\usepackage{algorithm2e}
\usepackage{amsmath}
\usepackage{amsthm}
\usepackage{amssymb}
\usepackage{graphicx}

\makeatletter

\providecommand{\tabularnewline}{\\}

\numberwithin{equation}{section}
\numberwithin{figure}{section}

\@ifundefined{showcaptionsetup}{}{%
 \PassOptionsToPackage{caption=false}{subfig}}
\usepackage{subfig}
\makeatother

\usepackage{babel}
\begin{document}

\title{A Conservative Discontinuous Galerkin Discretization for the Total
Energy Formulation of the Reacting Navier Stokes Equations}

\author{Ryan F. Johnson and Andrew D. Kercher}
\address{Laboratories for Computational Physics and Fluid Dynamics,  U.S. Naval Research Laboratory, 4555 Overlook Ave SW, Washington, DC 20375}

\begin{abstract}
This paper describes the total energy formulation of the compressible
reacting Navier-Stokes equations which is solved numerically using
a fully conservative discontinuous Galerkin finite element method
(DG). Previous applications of DG to the compressible reacting Navier-Stokes
equations required nonconservative fluxes or stabilization methods
in order to suppress unphysical oscillations in pressure that led
to the failure of simple test cases. In this paper, we demonstrate
that material interfaces with a temperature discontinuity result in
unphysical pressure oscillations if the species internal energy is
nonlinear with respect to temperature. We demonstrate that a temperature
discontinuity is the only type of material interface that results
in unphysical pressure oscillations for a conservative discretization
of the total energy formulation. Furthermore, we demonstrate that
unphysical pressure oscillations will be generated at any material
interface, including material interfaces where the temperature is
continuous, if the thermodynamics are frozen during the temporal integration
of the conserved state. Additionally, we demonstrate that the oscillations
are amplified if the specific heat at constant pressure is incorrectly
evaluated directly from the NASA polynomial expressions. Instead,
the mean value, which we derive in this manuscript, should be used
to compute the specific heat at constant pressure. This can reduce
the amplitude of, but not prevent, unphysical oscillations where the
species concentrations numerically mix. We then present solutions
to several test cases using a fully conservative DG discretization
of the total energy formulation. The test cases demonstrate that this
formulation does not generate spurious pressure oscillations for material
interfaces if the temperature is continuous and that it is better
behaved than frozen thermodynamic formulations if the temperature
is discontinuous. 
\end{abstract}
\maketitle
\global\long\def\middlebar{\,\middle|\,}%
\global\long\def\average#1{\left\{  \!\!\left\{  #1\right\}  \!\!\right\}  }%
\global\long\def\expnumber#1#2{{#1}\mathrm{e}{#2}}%

\section{Background\label{sec:Introduction}}

The discontinuous Galerkin finite element method (DG)~\citep{Bas97,Bas97_2,Coc98,Coc00,Arn02,Har02,Fid05,Luo06,Luo07,Luo08,Per08,Har13},
has been applied to the field of computational fluid dynamics with
great success over the past two decades. The method is fully conservative,
able to achieve high-order accuracy on unstructured grids, and has
a well-developed theory of adjoint consistency~\citep{Lu05,Har07,Har13,Har15},
which makes it a powerful tool for adjoint-based optimization. Furthermore,
an extension of DG, the Moving Discontinuous Galerkin Method with
Interface Condition Enforcement (MDG-ICE)~\citep{Cor17,Cor18}, maintains
high order accuracy for flows with interfaces and sharp gradients. 

Recent work has shown success in modeling combustion systems using
DG~\citep{Bil11,Lv15}. In these previous works, nonphysical pressure
oscillations were generated for continuous profiles of species and temperature 
with constant velocity and pressure. These oscillations have been attributed to variations in
the thermodynamic properties of multicomponent gases and it was previously
concluded that any fully conservative Godunov-type scheme would be
unable to maintain a pressure equilibrium across the material fronts~\citep{Abg88}.
This was shown conclusively for formulations that used variable ratio
of specific heats, $\gamma$, where $\gamma$ was a function of species
concentrations~\citep{Kar94,Abg96}. 

A nonconservative flux, referred to as the double flux method, was
implemented to avoid these oscillations in multicomponent reacting
flows~\citep{Abg01,Bil03,Bil11,Lv15}. The method assumes consistent
fluid thermal properties through a material interface, and thereby
breaks energy conservation to achieve the desired stability. This
method has been successfully applied to multidimensional reacting
flow problems, including detonations, and has been shown to be stable~\citep{Hou11,Lv15}.
Other methods, such as increasing the size of the state and solving
for additional transport equations, have also been employed to avoid
these unwanted oscillations in multicomponent flows~\citep{Joh12,Ter12,He17,Ma19,Ara19}.

In this work, we present the total energy formulation of the reacting
Navier-Stokes equations to simulate multicomponent flows that is suitable
for conservative DG discretizations. The formulation avoids unwanted
pressure oscillations without the need for nonconservative fluxes
or other stabilization methods by solving for temperature at each
degree of freedom such that the internal energy of the conserved fluid
is consistent with the mixture-averaged, temperature-based polynomial
expression for internal energy. This formulation is used to approximate
solutions to various test cases with material interfaces. The results
confirm that pressure oscillations are only generated at material
interfaces if the temperature is discontinuous. Additionally, we demonstrate
that unphysical pressure oscillations will be generated at any material
interface, including material interfaces where the temperature is
continuous, if the thermodynamics are frozen during the temporal integration
of the conserved state. Furthermore, we present verification for a
steady flame, demonstrating the ability of the method to model the
desired physics in combustion systems with high order accuracy without
generating pressure oscillations. 

\section{Formulations}

\subsection{The Total Energy Formulation\label{subsec:The-Total-Energy}}

The nonlinear conservation law, given in strong form, defined for
piecewise smooth, $\mathbb{R}^{m}$-valued functions $y$, and gradient
$\nabla y$, is

\begin{align}
\frac{\partial y}{\partial t}+\nabla\cdot\mathcal{F}\left(y,\nabla y\right)-\mathcal{S}\left(y\right)=0 & \textup{ in }\Omega,\label{eq:conservation-law-strong-form}\\
y\left(\cdot,t_{0}\right)-y_{0}=0 & \textup{ in }\Omega,\label{eq:conservation-law-initial-condition}\\
n\cdot\mathcal{F}\left(y,\nabla y\right)-n\cdot\mathcal{F}_{\Gamma}\left(y_{\Gamma},\nabla y\right)=0 & \textup{ in }\Gamma,\label{eq:conservation-law-boundary-condition}
\end{align}
for a given flux function $\mathcal{F}:\mathbb{R}^{m}\rightarrow\mathbb{R}^{m\times n}$
and source term $\mathcal{S}:\mathbb{R}^{m}\rightarrow\mathbb{R}^{m}$,
where $\Omega\subset\mathbb{R}^{n}$ is a given spatial domain and
$t$ denotes time. The initial conditions are given by $y_{0}$ in
Eq.~(\ref{eq:conservation-law-initial-condition}), while the boundary
conditions in Eq.~(\ref{eq:conservation-law-boundary-condition})
are imposed through the boundary flux, $\mathcal{F}_{\Gamma}\left(y_{\Gamma},\nabla y\right)=h_{\Gamma}\left(y\right)-\mathcal{F}_{\Gamma}^{\nu}\left(y_{\Gamma},\nabla y\right)$,
where $h_{\Gamma}\left(y\right)$ and $\mathcal{F}_{\Gamma}^{\nu}\left(y_{\Gamma},\nabla y\right)$
are the numerical and viscous fluxes, respectively, at the boundary.
The flux function 

\begin{equation}
F\left(y,\nabla y\right)=\left(\mathcal{F}^{c}\left(y\right)-\mathcal{F}^{v}\left(y,\nabla y\right)\right)\label{eq:flux_function}
\end{equation}
is defined in terms of the convective flux $\mathcal{F}^{c}\left(y\right)$,
which is only a function of the state $y$, and viscous flux $\mathcal{F}^{v}\left(y,\nabla y\right)$,
which is a function of the state and the gradient, $\nabla y$. The
reacting Navier-Stokes flow state variable is given by

\begin{equation}
y=\left(\rho v_{1},\ldots,\rho v_{n},\rho e_{t},C_{i},\ldots,C_{n_{s}}\right)\in\mathbb{R}^{m},\label{eq:reacting-navier-stokes-state}
\end{equation}
where $m=n+n_{s}+1$, $n$ is the number of spatial dimensions, $n_{s}$
is the number of thermally perfect species, $\rho$ is density, $\left(v_{1},\ldots,v_{n}\right)\in\mathbb{R}^{n}$
is velocity, $e_{t}$ is the specific total energy, and $C_{i}$ is
the concentration of species $i$. The density is calculated from
the concentrations as

\begin{equation}
\rho=\sum_{i=1}^{n_{s}}W_{i}C_{i},\label{eq:density_definition}
\end{equation}

\noindent where $W_{i}$ is the molecular weight of species $i$.

The $k$-th spatial convective flux component is given by
\begin{equation}
\mathcal{F}_{k}^{c}\left(y\right)=\left(\rho v_{k}v_{1}+p\delta_{k1},\ldots,\rho v_{k}v_{n}+p\delta_{kn},v_{k}\left(\rho e_{t}+p\right),v_{k}C_{1},\ldots,v_{k}C_{n_{s}}\right)\in\mathbb{R}^{m}.\label{eq:reacting-navier-stokes-spatial-convective-flux-component}
\end{equation}
The pressure is calculated from the equation of state,

\begin{equation}
p=R^{o}T\sum_{i=1}^{n_{s}}C_{i},\label{eq:EOS-1}
\end{equation}
where $T$ is the temperature and $R^{o}$ is the universal gas constant,
$8314.4621$ J/Kmol/K. The total energy in Eq.~(\ref{eq:reacting-navier-stokes-state})
is related to kinetic and internal energy by

\begin{equation}
\rho e_{t}=\rho u+\frac{1}{2}\sum_{k=1}^{n}\rho v_{k}v_{k},\label{eq:newton_eq}
\end{equation}
where $\rho u$ is a mass weighted sum of thermally perfect species
specific internal energies that are $n_{p}$-order polynomials with
respect to temperature, 

\begin{equation}
\rho u=\sum_{i=1}^{n_{s}}W_{i}C_{i}\sum_{k=0}^{n_{p}}a_{ik}T^{k}.\label{eq:internal_energy_polynomial}
\end{equation}

The $k$-th spatial component of the viscous flux is given by
\begin{equation}
\mathcal{F}_{k}^{v}\left(y,\nabla y\right)=\left(0,\tau_{1k},\ldots,\tau_{nk},\tau_{kj}v_{j}-W_{i}C_{i}h_{i}V_{ik}+\lambda\frac{\partial T}{\partial x_{k}},C_{1}V_{1k},\ldots,C_{n_{s}}V_{n_{s}k}\right)\in\mathbb{R}^{m},\label{eq:navier-stokes-viscous-flux-spatial-component}
\end{equation}

\noindent where $\lambda$ is the thermal conductivity, $\tau\in\mathbb{R}^{n\times n}$
is the viscous stress tensor, $h_{i}$ is the specific enthalpy of
species $i$, and $\left(V_{i1},\dots,V_{in}\right)\in\mathbb{R}^{n}$
is the diffusion velocity of species $i$. The $k$-th spatial component
of the viscous stress tensor is given by

\noindent 
\begin{equation}
\tau_{k}=\mu\left(\frac{\partial v_{1}}{\partial x_{k}}+\frac{\partial v_{k}}{\partial x_{1}}-\delta_{k1}\frac{2}{3}\sum_{j=1}^{n}\frac{\partial v_{j}}{\partial x_{j}},\ldots,\frac{\partial v_{n}}{\partial x_{k}}+\frac{\partial v_{k}}{\partial x_{n}}-\delta_{kn}\frac{2}{3}\sum_{j=1}^{n}\frac{\partial v_{j}}{\partial x_{j}}\right),\label{eq:reacting-navier-stokes-viscous-stress-tensor-component}
\end{equation}

\noindent where $\mu$ is the dynamic viscosity coefficient and $\left(x_{1},\ldots,x_{n}\right)\in\mathbb{R}^{n}$
are spatial coordinates. The transport properties are calculated using
mixture averaged properties. The species diffusion velocity is calculated
using the mixture averaged diffusion, 

\begin{equation}
V_{ik}=\frac{D_{i}}{C_{i}}\frac{\partial C_{i}}{\partial x_{k}}-\frac{D_{i}}{\rho}\frac{\partial\rho}{\partial x_{k}},\label{eq:diffusion_velocity}
\end{equation}
where $D_{i}$ is the mixture averaged diffusion coefficient of species
$i$ from~\citep{Kee89},
\begin{equation}
D_{i}=\frac{p_{atm}}{p\bar{W}}\frac{\sum_{j=1,j\ne i}^{n_{s}}X_{j}W_{j}}{\sum_{j=1,j\ne i}^{n_{s}}X_{j}/D_{ij}},\label{eq:diffusion}
\end{equation}
$p_{atm}=101325$ Pa, $X_{j}$ is the mole fraction of species $j$,
$D_{ij}$ is the diffusion coefficient of species $i$ to species
$j$, and $\bar{W}$ is the mixture molecular weight, $\bar{W}=\rho/\sum_{i=1}^{n_{s}}C_{i}$.
The mole fractions can be calculated directly from concentrations,
$X_{i}=C_{i}/\sum_{i=1}^{n_{s}}C_{i}$. The Wilke model~\citep{Wil50}
is used to calculate viscosity

\begin{eqnarray}
\mu & = & \sum_{i=1}^{n_{s}}\frac{X_{i}\mu_{i}}{X_{i}+\sum_{i=1,i\ne j}^{n_{s}}\left(X_{j}\phi_{ij}\right)},\label{eq:viscosity}
\end{eqnarray}
defined in terms of

\begin{eqnarray*}
\phi_{ij} & = & \frac{\left(1+\left(\frac{W_{j}}{W_{i}}\right)^{1/4}\sqrt{\left(\frac{\mu_{i}}{\mu_{j}}\right)}\right)^{2}}{\sqrt{8\left(1+\frac{W_{i}}{W_{j}}\right)}},
\end{eqnarray*}
where $\mu_{i}$ and $\mu_{j}$ are the species specific viscosities
for species $i$ and $j$ respectively. The Mathur model~\citep{Mat67}
is used to calculate conductivity,

\begin{equation}
\lambda=\frac{1}{2}\left(\sum_{i=1}^{n_{s}}X_{i}\lambda_{i}+\frac{1}{\sum_{i=1}^{n_{s}}\frac{X_{i}}{\lambda_{i}}}\right),\label{eq:conductivity}
\end{equation}
where $\lambda_{i}$ is the conductivity of species $i$.

Finally, the source term is given by

\begin{equation}
\mathcal{S}\left(y\right)=\left(0,\ldots,0,0,\omega_{1},\ldots,\omega_{n_{s}}\right)\in\mathbb{R}^{m},\label{eq:reacting-navier-stokes-source-term}
\end{equation}
where $\omega_{i}$ is the production rate of species $i$. The production
rate comes from the sum of the progress reaction rates from any arbitrary
number of reactions and reaction types. 

\subsection{The Ratio of Specific Heats Formulation\label{subsec:gamma_m_one_formulation}}

Here we present how the ratio of specific heats, $\gamma$, can be
used in the total energy formulation, described in Section~\ref{subsec:The-Total-Energy},
and we present the necessary constraints to keep the two formulations
consistent. We seek a formulation analogous to the calorically perfect
gas formulation, $\rho u=p/\left(\gamma-1\right)$. To do so we use
the definition of internal energy in terms of enthalpy and pressure,

\begin{equation}
\rho u=\rho h-p.\label{eq:internal_energy_enthalpy_relationship}
\end{equation}
Here the enthalpy is

\begin{equation}
\rho h=\rho\sum_{i=1}^{n_{s}}Y_{i}\int_{0}^{T}C_{p,i}dT=\rho\sum_{i=1}^{n_{s}}Y_{i}h_{i},\label{eq:total_enthalpy_definition}
\end{equation}
where $Y_{i}$ is the mass fraction of species $i$, $Y_{i}=W_{i}C_{i}/\rho$,
$C_{p,i}$ is the specific heat at constant pressure of species $i$,
and $h_{i}$ is the species specific enthalpy polynomial of temperature
that is degree $n_{p}$, $h_{i}=\sum_{k=0}^{n_{p}}a_{ik}T^{k}+R^{o}T$.
We reduce the definition of internal energy to achieve the equivalent
formulation that contains a similar expression to $\rho u=\frac{p}{\gamma-1}$
by introducing the mean value from reference temperature, $T_{0}$,
to current temperature, $T$, of $\bar{C}_{p,i}$ and $\bar{C}_{p}$
,

\begin{equation}
\bar{C}_{p,i}=\frac{1}{T-T_{0}}\int_{T_{0}}^{T}C_{p,i}dT=\frac{h_{i}-h_{i}^{0}}{T-T_{0}}\label{eq:cp_eff_enthalpy_specific}
\end{equation}
and

\begin{equation}
\bar{C}_{p}=\sum_{i=1}^{n_{s}}\frac{Y_{i}}{T-T_{0}}\int_{T_{0}}^{T}C_{p,i}dT=\frac{\sum_{i=1}^{n_{s}}Y_{i}\left(h_{i}-h_{i}^{0}\right)}{T-T_{0}}\label{eq:cp_eff_enthalpy}
\end{equation}
where $h_{i}^{0}$ is the species specific enthalpy at $T_{0}$. Using
the following definitions

\begin{eqnarray}
\bar{\gamma} & = & \frac{\bar{C}_{p}}{\bar{C}_{p}-R}=\frac{\sum_{i=1}^{n_{s}}\frac{Y_{i}\left(h_{i}-h_{i}^{0}\right)}{T-T_{0}}}{\sum_{i=1}^{n_{s}}\frac{Y_{i}\left(h_{i}-h_{i}^{0}\right)}{T-T_{0}}-R},\label{eq:cp_eff}\\
R & = & \frac{R^{o}\sum_{i=1}^{n_{s}}C_{i}}{\rho},\label{eq:R_mix}\\
\rho u & = & \frac{p}{\bar{\gamma}-1}+\rho\sum_{i=1}^{n_{s}}Y_{i}\left(h_{i}^{0}-\bar{C}_{p,i}T_{0}\right),\label{eq:internal_energy_gamma_m_one}
\end{eqnarray}
the inviscid total energy conservation without reactions becomes

\begin{eqnarray}
\frac{\partial\left(\frac{p}{\bar{\gamma}-1}+\frac{1}{2}\sum_{k=1}^{n}\rho v_{k}v_{k}\right)}{\partial t}+\nabla\cdot\left(\left(\frac{p}{\bar{\gamma}-1}+\frac{1}{2}\sum_{k=1}^{n}\rho v_{k}v_{k}+p\right)\left(v_{1},\dots,v_{n}\right)\right) & = & 0,\label{eq:gamma_m_one_equivalent}
\end{eqnarray}
where the term $\rho\sum_{i=1}^{n_{s}}Y_{i}\left(h_{i}^{0}-\bar{C}_{p,i}T_{0}\right)$
in Eq.~(\ref{eq:internal_energy_gamma_m_one}) is eliminated from
Eq.~(\ref{eq:gamma_m_one_equivalent}) by fixing $T_{0}$ to 0 K
and multiplying the non-reacting inviscid form of the species conservation
equations from Eq.~(\ref{eq:conservation-law-strong-form}) by $W_{i}h_{i}^{0}$
and summing over all species conservation equations. Eq.~(\ref{eq:gamma_m_one_equivalent})
is equivalent to the non-reacting inviscid form of the conservation
of energy from Eq.~(\ref{eq:conservation-law-strong-form}). Eq.~(\ref{eq:gamma_m_one_equivalent})
has the same form of the compressible Euler equations, and therefore
is convenient for routines that require the ratio of specific heats,
e.g., characteristic boundary conditions~\citep{Poi92}. However,
those routines require the flow to be non-reacting with constant thermodynamic
properties as $\bar{\gamma}$ is assumed to be constant.

\section{Material Discontinuities in Multicomponent Flows \label{sec:Discontinuities}}

A material discontinuity is defined as a discontinuity across which
there is no mass flow. The velocity and pressure are constant across
the discontinuity but other material quantities are not. In this section
we exact solutions for problems involving material interfaces by considering
the non-reacting inviscid formulation of Eqs.~(\ref{eq:conservation-law-strong-form})-(\ref{eq:conservation-law-boundary-condition})
where $\mathcal{F}^{v}\left(y,\nabla y\right)=\left(0,\dots,0\right)$
in Eq.~(\ref{eq:flux_function}) and $\mathcal{S}\left(y\right)=\left(0,\ldots,0\right)$
in Eq.~(\ref{eq:conservation-law-strong-form}). A discontinuous
solution, in one dimension satisfies, the inviscid form of Eqs.~(\ref{eq:conservation-law-strong-form})-(\ref{eq:conservation-law-boundary-condition})
if the jump in the flux is equal to the product of the jump in the
state and the material interface velocity~\citep{Mad84},

\begin{equation}
\mathcal{F}\left(y_{r}\right)-\mathcal{F}\left(y_{l}\right)=v_{s}\left(y_{r}-y_{l}\right),\label{eq:jump_condition_solution}
\end{equation}
where $y_{r}$ is the state on the right of the discontinuity, $y_{l}$
is the state on the left of the discontinuity, and $v_{s}$ is the
material velocity normal to the interface.

\begin{figure}[H]
\subfloat[\label{fig:interface_continuous}The speed of the material interface
is the slope of the trajectory $v_{s}=\bar{v}=\angle$.]{\begin{centering}
\includegraphics[width=0.45\columnwidth]{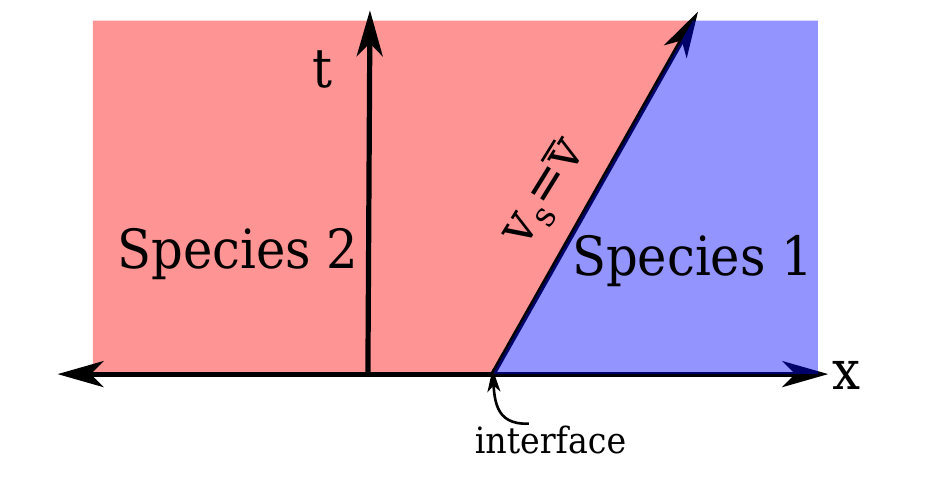}
\par\end{centering}
}\hfill{}\subfloat[\label{fig:interface_numerical}Numerical diffusion of material interface
at discrete times.]{\begin{centering}
\includegraphics[width=0.45\columnwidth]{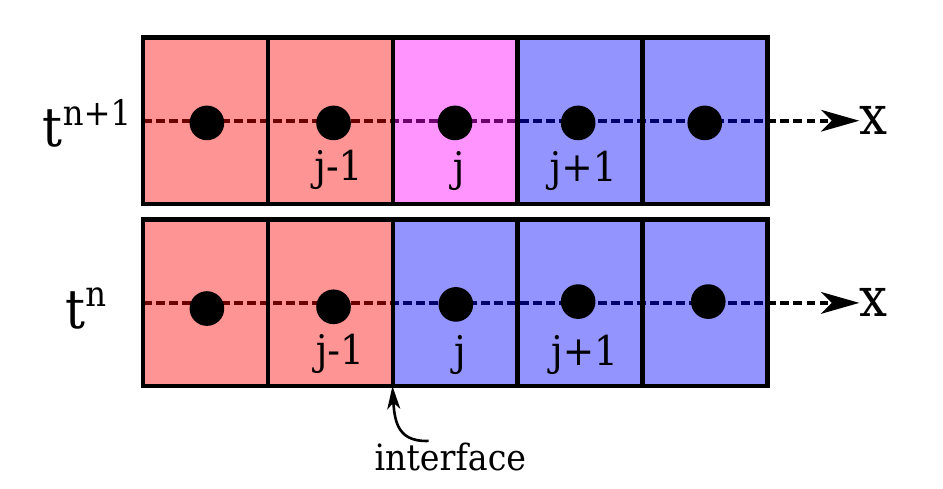}
\par\end{centering}
}
\centering{}\caption{\label{fig:interface_diagram}Diagrams of moving material interfaces.}
\end{figure}

Below we introduce a material discontinuity by considering a one-dimensional
two species discontinuity at $x_{j}$ where the velocity and pressure
are constant and the temperature is discontinuous,

\begin{eqnarray*}
v & = & \bar{v},\\
C_{1} & = & \begin{cases}
C_{1}^{0} & \text{if }x>x_{j}\\
0 & \text{otherwise}
\end{cases},\\
C_{2} & = & \begin{cases}
0 & \text{if }x<x_{j}\\
C_{2}^{0} & \text{otherwise}
\end{cases},\\
T & = & \begin{cases}
\eta\bar{T} & \text{if }x<x_{j}\\
\bar{T} & \text{otherwise}
\end{cases},\\
p & = & \bar{p.}
\end{eqnarray*}
The species with index $i=1$, species 1, has molecular weight $W_{1}$
and the species with index $i=2$, species 2, has molecular weight
$W_{2}$. The initial fluid state from Eq.~(\ref{eq:reacting-navier-stokes-state})
is therefore

\begin{equation}
y\left(x,t=0\right)=\begin{cases}
\left(W_{1}C_{1}^{0}\bar{v},\frac{1}{2}W_{1}C_{1}^{0}\bar{v}^{2}+W_{1}C_{1}^{0}\sum_{k=0}^{n_{p}}a_{1k}\bar{T}^{k},C_{1}^{0},0\right) & x>x_{j}\\
\left(W_{2}C_{2}^{0}\bar{v},\frac{1}{2}W_{2}C_{2}^{0}\bar{v}^{2}+W_{2}C_{2}^{0}\sum_{k=0}^{n_{p}}a_{2k}\left(\eta\bar{T}\right)^{k},0,C_{2}^{0}\right) & \text{otherwise}
\end{cases}.\label{eq:material_states}
\end{equation}
Substituting the fluid state from Eq.~(\ref{eq:material_states})
in Eq.~(\ref{eq:jump_condition_solution}) we arrive at the following
condition

\begin{eqnarray}
W_{1}C_{1}^{0}\bar{v}^{2}-W_{2}C_{2}^{0}\bar{v}^{2} & = & v_{s}\left(W_{1}C_{1}^{0}v-W_{2}C_{2}^{0}v\right),\label{eq:interface_momentum_sub}\\
\left(\frac{1}{2}W_{1}C_{1}^{0}v^{2}+C_{1}^{0}W_{1}\sum_{k=0}^{n_{p}}a_{1k}T^{k}+\bar{p}\right)\bar{v}\nonumber \\
-\left(\frac{1}{2}W_{2}C_{2}^{0}v^{2}+C_{2}^{0}W_{2}\sum_{k=0}^{n_{p}}a_{2k}\left(\eta T\right)^{k}+\bar{p}\right)\bar{v} & = & v_{s}\left(\frac{1}{2}W_{1}C_{1}^{0}v^{2}+W_{1}C_{1}^{0}\sum_{k=0}^{n_{p}}a_{1k}T^{k}\right.\nonumber \\
 &  & \left.-\frac{1}{2}W_{2}C_{2}^{0}v^{2}-W_{2}C_{2}^{0}\sum_{k=0}^{n_{p}}a_{2k}\left(\eta T\right)^{k}\right),\label{eq:interface_energy_sub}\\
C_{1}^{0}\bar{v} & = & v_{s}\left(C_{1}^{0}\right),\label{eq:interface_species_1_sub}\\
-C_{2}^{0}\bar{v} & = & v_{s}\left(-C_{2}^{0}\right).\label{eq:interface_species_2_sub}
\end{eqnarray}
Therefore a material discontinuity where velocity and pressure are
constant and the temperature is discontinuous satisfies Eqs.~(\ref{eq:conservation-law-strong-form})-(\ref{eq:conservation-law-boundary-condition})
with $v_{s}=\bar{v}$. A diagram of the space-time solution is shown
in Fig.~\ref{fig:interface_continuous}. 

We now present the effect of a linear discretization on the same two
species discontinuity. Using the notation from Abgrall and Karni~\citep{Abg01},
the inviscid non-reacting conservation equations can be written as

\begin{align}
\delta\left(\rho v\right)+\nu\Delta\left(\rho v^{2}+p\right)= & 0,\label{eq:linear_momentum}\\
\delta\left(\rho e_{t}\right)+\nu\Delta\left(\left(\rho e_{t}+p\right)v\right)= & 0,\label{eq:linear_energy}\\
\delta\left(C_{i}\right)+\nu\Delta\left(C_{i}v\right)= & 0\text{ for }i=1\dots n_{s},\label{eq:linear_species}
\end{align}
where the inviscid forms of Eqs.~(\ref{eq:conservation-law-strong-form})-(\ref{eq:conservation-law-boundary-condition})
have been linearized with respect to space and time. Here, $\delta\left(\right)=\left(\right)_{j}^{n+1}-\left(\right)_{j}^{n}$
denotes the temporal change of the state, $\Delta\left(\right)$ denotes
spatial variation across the interface $\Delta\left(\right)=\left(\right)_{j}^{n}-\left(\right)_{j-1}^{n}$,
and $\nu=\frac{\Delta t}{\Delta x}$ where $\Delta t$ is the chosen
time step and $\Delta x$ is the spatial distance across the interface.
The material interface is initially between two nodes, $j$ and $j-1$,
as depicted at time $t^{n}$ in Fig.~\ref{fig:interface_numerical}.
Specifically, the initial flow state at $t^{n}$ is

\begin{eqnarray}
y_{j}^{n} & = & \left(W_{1}C_{1}^{0}\bar{v},\frac{1}{2}W_{1}C_{1}^{0}\bar{v}^{2}+W_{1}C_{1}^{0}\sum_{k=0}^{n_{p}}a_{1k}\bar{T}^{k},C_{1}^{0},0\right),\label{eq:initial_state_numerical_1}\\
y_{j-1}^{n} & = & \left(W_{2}C_{2}^{0}\bar{v},\frac{1}{2}W_{2}C_{2}^{0}\bar{v}^{2}+W_{2}C_{2}^{0}\sum_{k=0}^{n_{p}}a_{2k}\left(\eta\bar{T}\right)^{k},0,C_{2}^{0}\right).\label{eq:initial_state_numerical_2}
\end{eqnarray}
For simplification purposes, we define the initial concentration of
species 2 in terms of the initial concentration of species 1 through
the constant initial pressure conditions, $p_{j}^{n}=p_{j-1}^{n}=\bar{p}$,
and the equation of state, Eq.~(\ref{eq:EOS-1}),

\begin{equation}
R^{o}\bar{T}C_{1}^{0}=R^{o}\eta\bar{T}C_{2}^{0}\rightarrow C_{2}^{0}=\frac{C_{1}^{0}}{\eta}.\label{eq:species_relationship}
\end{equation}
The species conservation, Eq.~(\ref{eq:linear_species}), gives the
concentrations at $t^{n+1}$ in terms of the initial species 1 concentration,

\begin{eqnarray}
C_{1,j}^{n+1} & = & C_{1}^{0}-\nu\bar{v}C_{1}^{0},\label{eq:change_species_1}\\
C_{2,j}^{n+1} & = & \nu\bar{v}C_{2}^{0}=\nu\bar{v}\frac{C_{1}^{0}}{\eta}.\label{eq:change_species_2}
\end{eqnarray}
Eqs.~(\ref{eq:change_species_1})~and~(\ref{eq:change_species_2})
show that there is numerical mixing of the species at time $t^{n+1}$
and node $j$, as depicted in Fig.~\ref{fig:interface_numerical}.
This is a departure from the exact solution that satisfies the interface
condition, depicted in Fig.~\ref{fig:interface_continuous}, and
we continue in this section by examining the effect that the numerical
mixing of the species concentrations has on the stability of the material
interface.

Substituting the concentrations from Eqs.~(\ref{eq:change_species_1})~and~(\ref{eq:change_species_2})
into Eq.~(\ref{eq:density_definition}) gives the density at $t^{n+1}$,

\begin{equation}
\rho_{j}^{n+1}=\left(W_{1}+\nu\bar{v}\left(\frac{W_{2}}{\eta}-W_{1}\right)\right)C_{1}^{0}.\label{eq:change_density}
\end{equation}
Using Eqs.~(\ref{eq:change_species_1})-(\ref{eq:change_density})
and Eq.~(\ref{eq:density_definition}) for the initial density and
substituting into Eq.~(\ref{eq:linear_momentum}) reveals that the
velocity remains constant,

\begin{eqnarray}
\left(W_{1}+\nu\bar{v}\left(\frac{W_{2}}{\eta}-W_{1}\right)\right)C_{1}^{0}v^{n+1}-W_{1}C_{1}^{0}\bar{v}+\nu\bar{v}^{2}\left(W_{1}C_{1}^{0}-W_{2}\frac{C_{1}^{0}}{\eta}\right) & = & 0\rightarrow v_{j}^{n+1}=\bar{v}.\label{eq:constant_velocity}
\end{eqnarray}

Using Eqs.~(\ref{eq:change_species_1})-(\ref{eq:constant_velocity})
we consider the change in total energy to analyze the stability of
the material interface. We derive a relationship for kinetic energy
by multiplying Eq.~(\ref{eq:linear_momentum}) by $\frac{1}{2}\bar{v}$,

\begin{eqnarray}
\left(\frac{1}{2}\rho_{j}^{n+1}\bar{v}^{2}-\frac{1}{2}\rho_{j}^{n}\bar{v}^{2}+\nu\bar{v}\left(\frac{1}{2}\rho_{j}^{n}\bar{v}^{2}-\frac{1}{2}\rho_{j-1}^{n}\bar{v}^{2}\right)\right) & = & 0,\label{eq:kinetic_energy_eliminate}
\end{eqnarray}
 and we derive a relationship for pressure by noting $pv$ is constant
across the interface at $t^{n}$,

\begin{eqnarray}
\Delta\left(pv\right) & = & 0.\label{eq:pressure_differential_eliminate}
\end{eqnarray}
Combining Eq.~(\ref{eq:kinetic_energy_eliminate}) and Eq.~(\ref{eq:pressure_differential_eliminate})
with Eq.~(\ref{eq:linear_energy}) we remove the kinetic energy,
contained in $\rho e_{t}=\rho v^{2}/2+\rho u$, and the pressure term
to yield a linear relationship for the internal energy across the
interface,

\begin{eqnarray}
\delta\rho u+\nu v\Delta\left(\rho u\right) & = & 0.\label{eq:linear_internal_energy}
\end{eqnarray}
We substitute Eq.~(\ref{eq:internal_energy_polynomial}), Eq.~(\ref{eq:species_relationship}),
and Eqs.~(\ref{eq:change_species_1})-(\ref{eq:change_density})
in Eq.~(\ref{eq:linear_internal_energy}) and arrive at $n_{p}$
expressions for the temperature at time $t^{n+1}$ by collecting like
terms,

\begin{eqnarray}
T_{j}^{n+1} & = & \bar{T}\frac{\left(W_{1}a_{11}-\nu\bar{v}\left(W_{1}a_{11}-W_{2}a_{21}\right)\right)}{W_{1}\left(1-\nu\bar{v}\right)a_{11}+\frac{W_{2}}{\eta}\left(\nu\bar{v}\right)a_{21}}\nonumber \\
\vdots & \vdots & \vdots\nonumber \\
\left(T_{j}^{n+1}\right)^{n_{p}} & = & \left(\bar{T}\right)^{n_{p}}\frac{\left(W_{1}a_{1n_{p}}-\nu\bar{v}\left(W_{1}a_{1N_{p}}-W_{2}a_{2N_{p}}\eta^{n_{p}-1}\right)\right)}{W_{1}\left(1-\nu\bar{v}\right)a_{1n_{p}}+\frac{W_{2}}{\eta}\left(\nu\bar{v}\right)a_{2n_{p}}}.\label{eq:temperature_relationships}
\end{eqnarray}
Finally, the change in pressure is given as

\begin{eqnarray}
p_{j}^{n+1}-p_{j}^{n} & = & R^{o}T_{j}^{n+1}\left(C_{1}^{0}-\nu\bar{v}C_{1}^{0}+\nu\bar{v}\frac{C_{1}^{0}}{\eta}\right)-R^{o}\bar{T}C_{1}^{0}.\label{eq:pressure_change}
\end{eqnarray}

From analyzing Eqs.~(\ref{eq:temperature_relationships})~and~(\ref{eq:pressure_change})
we come to similar conclusions to those of Jenny et al~\citep{Jen97},
that pressure oscillations, $p_{j}^{n+1}-p_{j}^{n}\ne0$, do not exist
if one of the following conditions is true
\begin{enumerate}
\item The temperature is continuous, $\eta=1$.
\item The contact discontinuity remains grid aligned, $\nu\bar{v}=1$.
\item The contact discontinuity is stationary, $\bar{v}=0$.
\item The internal energies are linear, $n_{p}=1$, with respect to temperature
\emph{and} the species are the same across the interface, i.e., molecular
weights are constant across the interface, $W_{1}=W_{2}$, and the
internal energies are the same across the interface, $a_{1k}=a_{2k}$.
\end{enumerate}
For condition (1), the numerical mixing of species concentrations,
Eqs.~(\ref{eq:change_species_1})~and~(\ref{eq:change_species_2}),
inside the cell does not cause a pressure oscillation as both species
are at the same temperature despite having different internal energies. 

When $\eta\ne1$ the temperature is discontinuous and stabilization,
e.g., artificial viscosity, would be required if (2)-(4) were not
satisfied. Satisfaction of condition (2) would requires an interface
fitting method~\citep{Cor17,Cor18,Zah17,Zah18} that dynamically
fits a priori unknown discontinuities and is therefore beyond the
scope of this manuscript. Condition (3) is a trivial case. Condition
(4) applies to ideal gases that have a linear relationship between
temperature and internal energy and are assumed to be the same species
in all regions of the flow.

Applying the same linearization to Eq.~(\ref{eq:gamma_m_one_equivalent})
we can arrive at a similar relationship for the internal energy based
on $\bar{\ensuremath{\gamma}}$ and $p$,

\begin{eqnarray}
\frac{p_{j}^{n+1}}{\bar{\gamma}_{j}^{n+1}-1} & = & \frac{\bar{p}}{\bar{\gamma_{1}}-1}-\nu v\left(\frac{\bar{p}}{\bar{\gamma_{1}}-1}-\frac{\bar{p}}{\bar{\gamma}_{2}-1}\right),\label{eq:pressure_gamma_relationship}
\end{eqnarray}
where $\bar{\gamma}_{1}$ and $\bar{\gamma}_{2}$ are the known specific
heat ratios of the right and left hand side based on $C_{1}^{0}$
at temperature $\bar{T}$ and $C_{2}^{0}$ at temperature $\eta\bar{T}$,
respectively. The equivalent process for the ratio of specific heats
formulation would be to use the definition of pressure and $\bar{\gamma}$
in terms of known concentrations, $C_{1}^{n+1}$, $C_{2}^{n+1}$,
$C_{i}^{0}$, and $C_{2}^{0}$, and temperatures, $\eta\bar{T}$ and
$\bar{T}$, to solve for $T^{n+1}$. This results in similar nonlinear
relationships for temperature but instead from the $h_{i}$ polynomials.
It follows that the same stability properties found for the total
energy formulation apply to the ratio of specific heats formulation.

\section{Numerical Methods}

In order to discretize the Eqs.~(\ref{eq:conservation-law-strong-form})-(\ref{eq:conservation-law-boundary-condition}),
the DG method assumes that $\Omega$ can be subdivided into a mesh
$\mathcal{T}_{h}=\left\{ \kappa\right\} $, consisting of disjoint
cells $\kappa$ such that $\bar{\Omega}=\cup_{\kappa\in\mathcal{T}_{h}}\bar{\kappa}$.
We also consider $\mathcal{I}_{h}$, consisting of interior interfaces
defined between pairs of cells, so that for each $\epsilon\in\mathcal{I}_{h}$,
there exists a pair of $\kappa^{+},\kappa^{-}\in\mathcal{T}_{h}$
such that $\epsilon=\kappa^{+}\cap\kappa^{-}$. A discrete subspace,
consisting of piecewise polynomials of degree $p$, is defined over
$\mathcal{T}_{h}$,
\begin{equation}
V_{h}=\left\{ v\mid v\in P^{p}\left(\kappa\right)\text{ for }\kappa\in\mathcal{T}_{h}\left(\Omega\right)\right\} .\label{eq:discrete-subspace}
\end{equation}
From this, a DG (semi-)discretization is obtained, find $\frac{\partial y}{\partial t}\in V_{h}$
such that

\begin{gather}
\sum_{\kappa\in\mathcal{T}_{h}}\int_{\kappa}\frac{\partial y}{\partial t}v+\int_{\partial\kappa}\left(h\left(y^{+},y^{-},n\right)\right)v^{+}-\int_{\partial\kappa}\average{\mathcal{F}^{\nu}\left(y,\nabla y\right)}\cdot nv^{+}\qquad\qquad\qquad\qquad\nonumber \\
\qquad\qquad\qquad-\int_{\partial\kappa}\delta\left(y^{+},y^{-},n\right)v^{+}-\int_{\kappa}\mathcal{F}\left(y,\nabla y\right)\cdot\nabla v-\int_{\kappa}\mathcal{S}\left(y\right)\cdot v=0\qquad\forall v\in V_{h},\label{eq:semi-discretization}
\end{gather}
where $h\left(y^{+},y^{-},n\right)$ is the numerical flux, choose
to be the HLLC approximate Riemann~\citep{Tor13}, see also Appendix
B of ~\citep{Lv15},

\[
\average{\mathcal{F}^{\nu}\left(y,\nabla y\right)}=\frac{1}{2}\left(\mathcal{F}^{\nu}\left(y^{+},\nabla y^{+}\right)+\mathcal{F}^{\nu}\left(y^{-},\nabla y^{-}\right)\right),
\]
is the average viscous flux, and $\delta\left(y^{+},y^{-},n\right)$
is a penalty term that is required for stability and is implemented
via the BR2 formulation~\citep{Bas98,Bas00,Bas02}. On the exterior
interfaces, $\partial\kappa\notin\mathcal{I}_{h}$, the numerical
flux is defined consistently with the imposed boundary condition Eq.~(\ref{eq:conservation-law-boundary-condition})
and 
\[
\average{\mathcal{F}^{\nu}\left(y,\nabla y\right)}=\left(\mathcal{F}^{\nu}\left(y_{\Gamma},\nabla y^{+}\right)\right).
\]
The DG space semi-discretization is integrated temporally with a strong-stability-preserving
Runge-Kutta method (SSP-RK2)~\citep{Got01}. This discretization
of convection and diffusion is combined with a stiff ODE solver for
the reaction term via Strang splitting~\citep{Str68}.

\subsection{Implementation of Thermodynamics}

\begin{figure}[H]
\begin{centering}
\includegraphics[width=0.6\columnwidth]{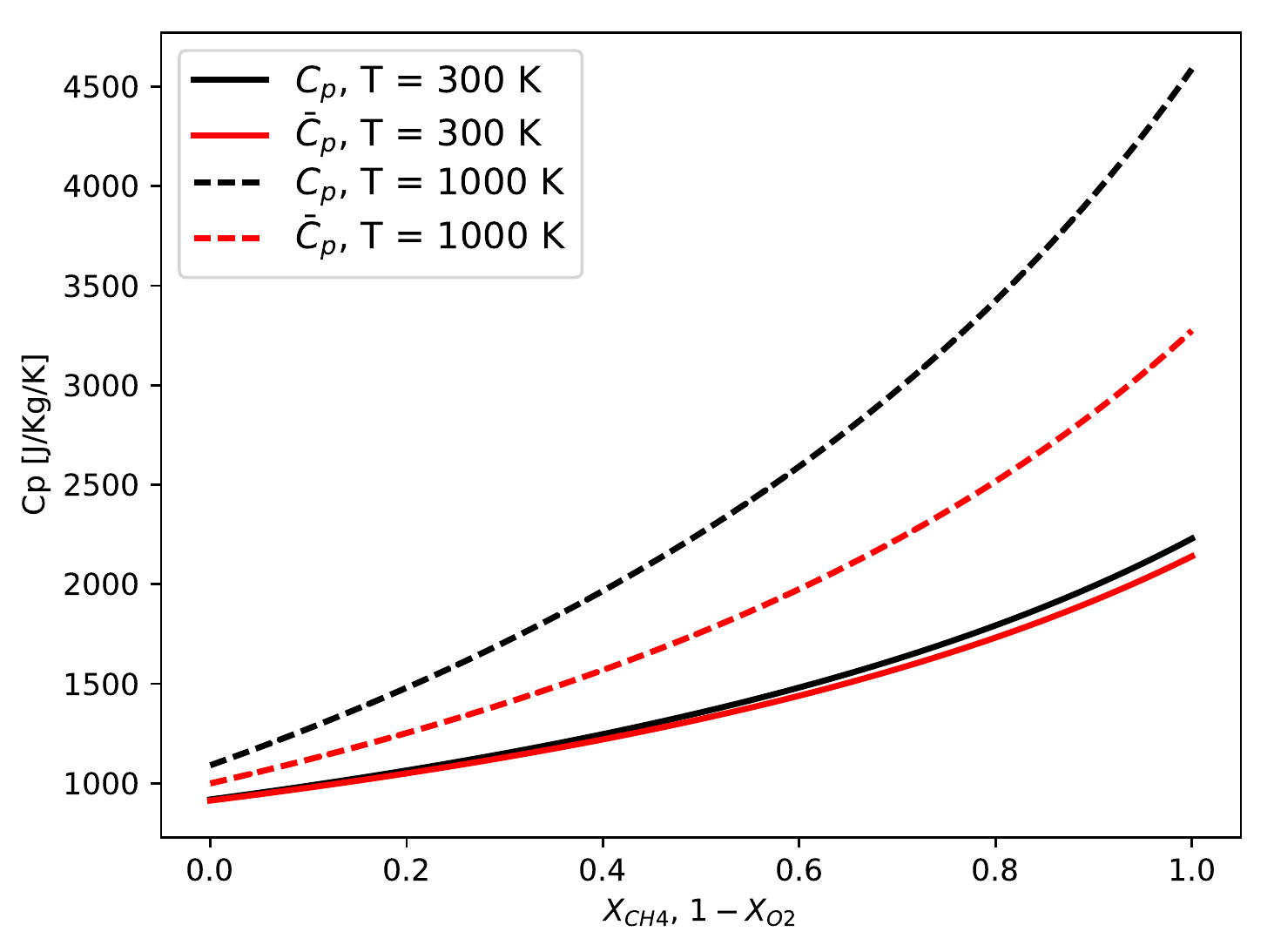}
\par\end{centering}
\caption{The difference between $\bar{C_{p}}$ and $C_{p}$ with reference
temperature $T_{0}=200$ K, for a varying mixture of CH4 and O2 at
two different temperatures, $T=300$ K and $T=1000$K.\label{fig:Cp_difference}}
\end{figure}

It is important to note that $\bar{C}_{p,i}$, Eq.~(\ref{eq:cp_eff_enthalpy_specific}),
and $\bar{C_{p}}$, Eq.~(\ref{eq:cp_eff_enthalpy}), from Section~\ref{subsec:gamma_m_one_formulation}
are not equivalent to the evaluations of $C_{p,i}$ and $C_{p}$ from
polynomial expressions, e.g., the analytic form of NASA’s polynomial
representations~\citep{Mcb02}. Rather, $\bar{C}_{p,i}$ and $\bar{C}_{p}$
can be viewed as the mean value of $C_{p,i}$ and $C_{p}$, where
$\bar{C_{p}}$ approaches $C_{p}$ as $T-T_{o}\rightarrow0$. This
difference can be shown mathematically by using the following polynomial
definitions of specific heat at constant pressure,

\begin{eqnarray}
C_{pi} & = & \sum_{k=0}^{n_{p}}b_{ik}T^{k}\label{eq:species_cp_polynomail}
\end{eqnarray}
and mixture averaged $C_{p}$, 

\begin{eqnarray}
C_{p} & = & \sum_{i=1}^{n_{s}}Y_{i}\sum_{k=0}^{n_{p}}b_{ik}T^{k}.\label{eq:mixture_average_cp_polynomial}
\end{eqnarray}
We arrive at the total enthalpy in polynomial form by integrating
of Eqs.~(\ref{eq:species_cp_polynomail})~and~(\ref{eq:mixture_average_cp_polynomial})
from $0$ to $T$ and substituting the result into Eq.~(\ref{eq:total_enthalpy_definition}),

\begin{eqnarray}
\rho h & =\rho\sum_{i=1}^{n_{s}}Y_{i}\left(\int_{0}^{T}C_{p,i}dT\right)= & \rho\sum_{i=1}^{n_{s}}Y_{i}\left(\sum_{k=0}^{n_{p}}b_{ik}\frac{T^{k+1}}{k+1}\right).\label{eq:total_enthalpy_cp_polynomial}
\end{eqnarray}
By substituting Eq.~(\ref{eq:total_enthalpy_cp_polynomial}) in Eq.~(\ref{eq:cp_eff_enthalpy}),
we arrive at $\bar{C_{p}}$ in terms of the $C_{pi}$ polynomial coefficients
and temperature,

\begin{equation}
\bar{C_{p}}=\frac{\sum_{i=1}^{n_{s}}Y_{i}\left(\sum_{k=0}^{n_{p}}b_{ik}\frac{T^{k+1}}{k+1}-\sum_{k=0}^{n_{p}}b_{ik}\frac{T_{0}^{k+1}}{k+1}\right)}{T-T_{0}}.
\end{equation}
Therefore, $\bar{C_{p}}$ and $C_{p}$ are only equivalent if $C_{p}$
is constant with respect to temperature, i.e., $n_{p}=0$. Fig.~\ref{fig:Cp_difference}
shows the difference between $C_{p}$ evaluated from NASA polynomials
and $\bar{C_{p}}$ evaluated from Eq.~(\ref{eq:cp_eff_enthalpy})
using a reference temperature, $T_{o}$, of 200 K and mixture of methane,
$CH_{4}$, and oxygen, $O_{2}$. Both $\bar{C_{p}}$ and $C_{P}$
were evaluated at a temperature, $T$, of 300 K and 1000 K with the
mixture varying from pure methane to pure oxygen, $X_{CH4}=1-X_{O2}$.
The values for $C_{p}$ and $\bar{C_{p}}$ are significantly different
at higher temperatures due to the large difference between the actual
temperature and the reference temperature, $T-T_{0}$. 

In Section~\ref{sec:test_cases} three different methods are used
to evaluate the thermodynamics in the convective flux, $\mathcal{F}_{k}^{c}\left(y\right)$,
for various test cases. Table~\ref{tab:flux_evaluation} shows the
steps required to calculate the convective fluxes. The first column
corresponds to the total energy formulation that requires the solution
of a nonlinear equation for the temperature. The solution ensures
equivalency between the computed internal energy and the internal
energy given by the evaluation of the corresponding polynomial expression,
i.e., find $T$ such that

\begin{equation}
0=\rho u-\sum_{i=1}^{n_{s}}W_{i}C_{i}\sum_{k=1}^{K}a_{ik}T^{k},
\end{equation}
where the internal energy, $\rho u=\rho e_{t}-\frac{1}{2}\sum_{k=1}^{n}\rho v_{k}v_{k}$,
and concentrations, $C_{i}$, are computed from the known state, $y$.
In this work the temperature is computed such that the following is
satisfied to machine precision for a given an initial temperature:

\begin{equation}
\delta T=\frac{\rho u-\sum_{i=1}^{n_{s}}W_{i}C_{i}\sum_{k=1}^{K}a_{ik}T^{k}}{\frac{\partial\rho u}{\partial T}},
\end{equation}
where $\delta T$ is the temperature decrement corresponding to Newton's
method and

\begin{equation}
\frac{\partial\rho u}{\partial T}=\sum_{i=1}^{n_{s}}W_{i}C_{i}\sum_{k=1}^{K}ka_{ik}T^{k-1}
\end{equation}
is the partial derivative of internal energy with respect to temperature.
In practice, we observe that the temperature converges within five
nonlinear iterations. Once the temperature and state are known, pressure
is computed by evaluating Eq.~(\ref{eq:EOS-1}). The second column
of Table~\ref{tab:flux_evaluation} corresponds to a frozen thermodynamics
evaluation of convective flux where the ratio of specific heats formulation
is used and $\bar{\gamma}$ and $\bar{C_{p}}$ are held constant throughout
each time step. The third row for the frozen thermodynamics is split
into two options. The left corresponds to using $\bar{C_{p}}$ in
the evaluation of the flux and the right corresponds to using $C_{p}$,
which is evaluated from NASA polynomials expressions.

\begin{table}[H]
\caption{Procedures used to evaluate the convective flux\label{tab:flux_evaluation}.}

\centering{}%
\begin{tabular}{|c|>{\centering}p{1.75in}|>{\centering}p{1.75in}|>{\centering}p{1.75in}|}
\hline 
Step & Total Energy Formulation & \multicolumn{2}{c|}{Ratio of Specific Heats Formulation with Frozen Thermodynamics}\tabularnewline
\hline 
\hline 
1 & Calculate temperature by solving the nonlinear Eq.~(\ref{eq:newton_eq}). & \multicolumn{2}{c|}{Use frozen $\bar{\gamma}$ and $\bar{C_{p}}$ to calculate pressure}\tabularnewline
\hline 
2 & Calculate pressure from temperature via Eq.~(\ref{eq:EOS-1})  & \multicolumn{2}{c|}{Calculate $h_{HLLC}\left(y^{+},y^{-},n\right)$ and $\mathcal{F}_{k}^{c}\left(y\right)$}\tabularnewline
\hline 
3 & Calculate $h_{HLLC}\left(y^{+},y^{-},n\right)$ and $\mathcal{F}_{k}^{c}\left(y\right)$ & Calculate and freeze new $\bar{C_{p}}$ from Eq.~ (\ref{eq:cp_eff_enthalpy})
and freeze  & Calculate and freeze new $\bar{C_{p}}$ as equivalent to $C_{p}$
from Eq. ~(\ref{eq:mixture_average_cp_polynomial}) and freeze \tabularnewline
\hline 
4 &  & \multicolumn{2}{c|}{Calculate and freeze new $\bar{\gamma}$ from $\bar{C_{p}}$ and $R$}\tabularnewline
\hline 
\end{tabular}
\end{table}
If $\bar{\gamma}$ is frozen as presented in the second column in
Table~\ref{tab:flux_evaluation} then $\gamma_{j}^{n+1}$ in Eq.~(\ref{eq:pressure_gamma_relationship})
is $\bar{\gamma}_{1}$. A pressure oscillation is now generated regardless
of the type of material interface, 

\begin{equation}
p_{j}^{n+1}=\bar{p}\left(1-\nu v\left(1-\frac{\bar{\gamma_{1}}-1}{\bar{\gamma}_{2}-1}\right)\right).\label{eq:freeze_gamma_pressure_oscillation}
\end{equation}
The freezing of thermodynamics would therefore require other methods
to identify and stabilize the nonphysical pressure oscillations even
in smooth regions of the flow where $\eta=1$. Previous work showed
that pressure oscillations develop even in the presence of continuous
temperature profiles which was due to the freezing of thermodynamics
\citep{Bil03,Bil11,Hou11,Lv15,Ara19}. In the following section we
explore the magnitude of these pressure oscillations due to frozen
thermodynamics and compare to the behavior of the total energy formulation. 

\section{Test Cases and Verification\label{sec:test_cases}}

In this section we solve several test cases with the total energy
formulation and verify the analysis from Section~\ref{sec:Discontinuities}.
Furthermore, we compare the stability of the total energy formulation
to the ratio of specific heats formulation using both frozen $C_{p}$
and frozen $\bar{C_{p}}$. We then verify the total energy formulation
by comparing to a one-dimensional premixed hydrogen flame simulation
generated by Cantera's free-flame solver~\citep{cantera}. In the
test cases below, the time step is restricted by the Courant-Friedrichs-Lewy
number, CFL, defined as

\begin{equation}
\mathrm{CFL}=\frac{\Delta t}{\left(2p+1\right)\min\left(\Delta x\right)}\left(||v||+c\right),
\end{equation}
where $p$ is the polynomial degree and $c$ is the speed of sound,
defined as $c=\sqrt{\gamma RT}$. Here $\gamma$ is the ratio of specific
heats which is defined as $\gamma=\frac{C_{p}}{C_{p}-R}$ for both
the total energy formulation and the ratio of specific heats formulation
with frozen $C_{p}$ and $\gamma=\bar{\gamma}$ for the ratio of specific
heats formulation with frozen $\bar{C_{p}}$. 

\subsection{Discontinuities\label{subsec:Discontinuities}}

We approximate exact solutions corresponding to the discontinuities
described in Section~\ref{sec:Discontinuities} by solving the non-reacting,
inviscid formulation of the Navier-Stokes Eqs.~(\ref{eq:conservation-law-strong-form})-(\ref{eq:conservation-law-boundary-condition})
using $\mathrm{DG}(p=1)$, $\mathrm{DG}(p=2)$, and $\mathrm{DG}(p=3)$,
without artificial viscosity or limiting. The test problems are run
on a periodic domain for one full cycle with $\mathrm{CFL}=0.1$.
For each test case we constructed two fictitious species, $n_{s}=2$,
with different molecular weights, $W_{1}=20$ and $W_{2}=70$. Here
we used a nonlinear function for internal energy, with $n_{p}=3$,
$a_{1k}=\{0,2.08e\times10^{4},83.1,-2.77\times10^{-2}\}$ and $a_{2k}=\{0,3.33e\times10^{4},83.1,-2.22\times10^{-2}\}$,
of the form $u_{i}=\sum_{k=0}^{n_{p}}a_{ik}T^{k}$ where $u_{i}$
is the units J/kg with temperature, $T$, in K. The enthalpy of each
species is therefore $h_{i}=\sum_{k=0}^{n_{p}}a_{ik}T^{k}+R^{o}T$
and the specific heat at constant pressure of each species is $C_{p,i}=\sum_{k=0}^{n_{p}}ka_{ik}T^{k-1}+R^{o}$.
For each test case the domain is $0.1$ m, $\left\{ x:x\in\left(0,0.1\right)\right\} $,
with grid spacing of $h=0.002$ m. The test cases are solved using
both the total energy formulation and the ratio of specific heats
formulation with frozen thermodynamics as outlined in Table~\ref{tab:flux_evaluation}. 

\subsubsection{Species Discontinuities at Constant Temperature\label{subsec:species-discontinuity-constant-temperature}}

\begin{figure}[H]
\subfloat[$\mathrm{DG}\left(p=1\right)$ solution \label{fig:species_discontinuity_constT_nl_p1}]{\includegraphics[width=0.32\columnwidth]{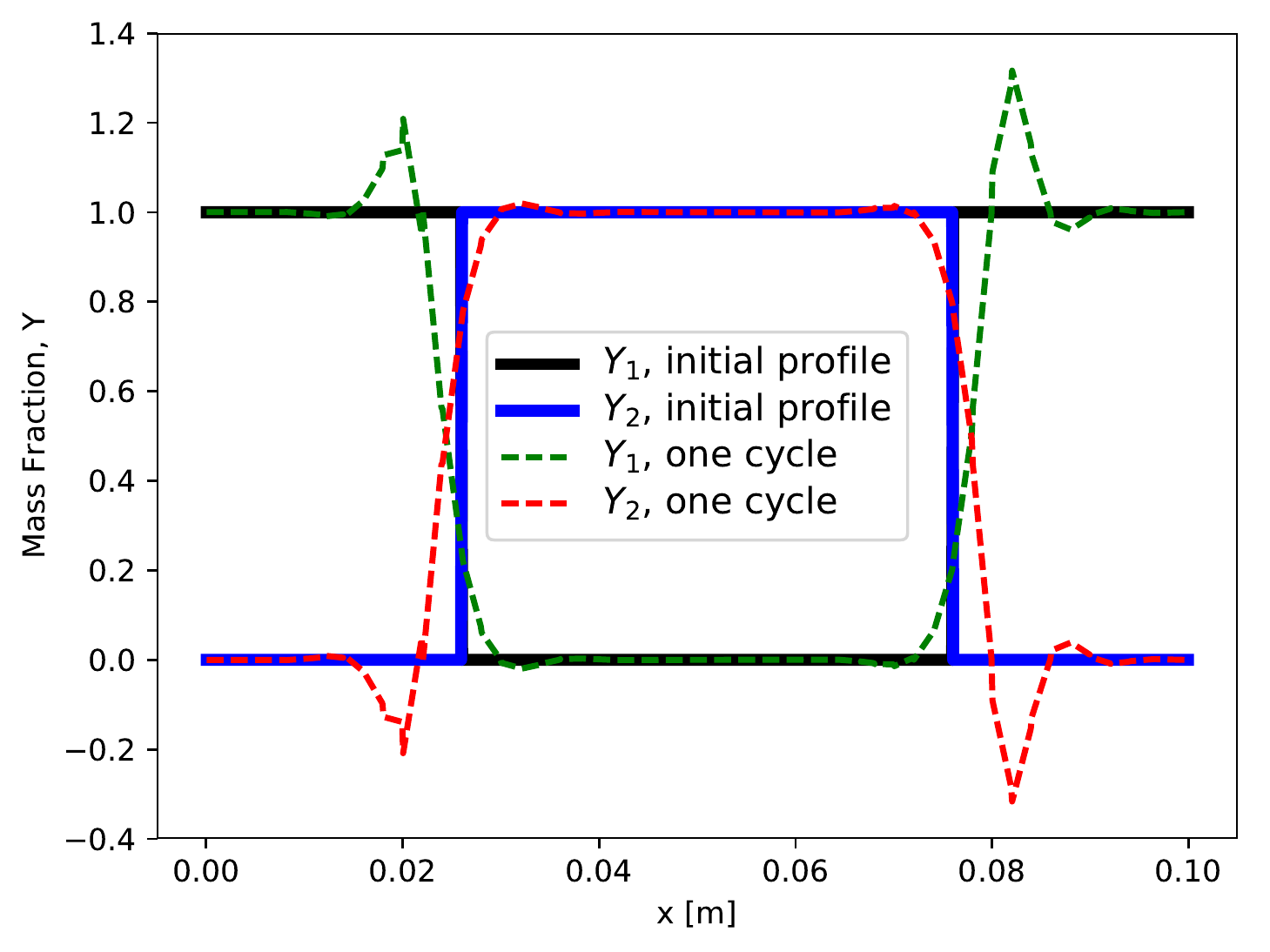}

}\hfill{}\subfloat[$\mathrm{DG}\left(p=2\right)$ solution \label{fig:species_discontinuity_constT_nl_p2}]{\includegraphics[width=0.32\textwidth]{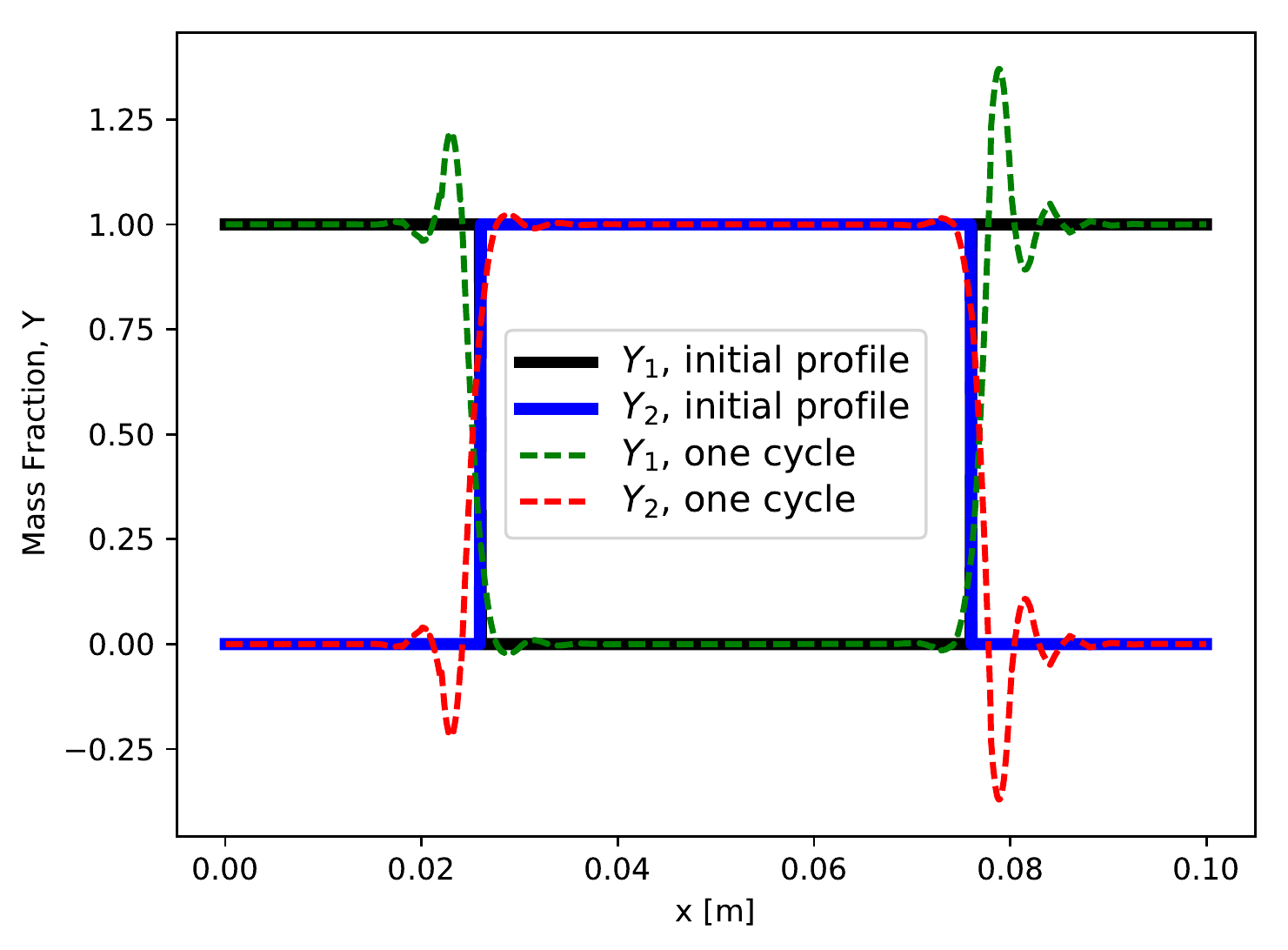}

}\hfill{}\subfloat[$\mathrm{DG}\left(p=3\right)$ solution \label{fig:species_discontinuity_constT_nl_p3}]{\includegraphics[width=0.32\textwidth]{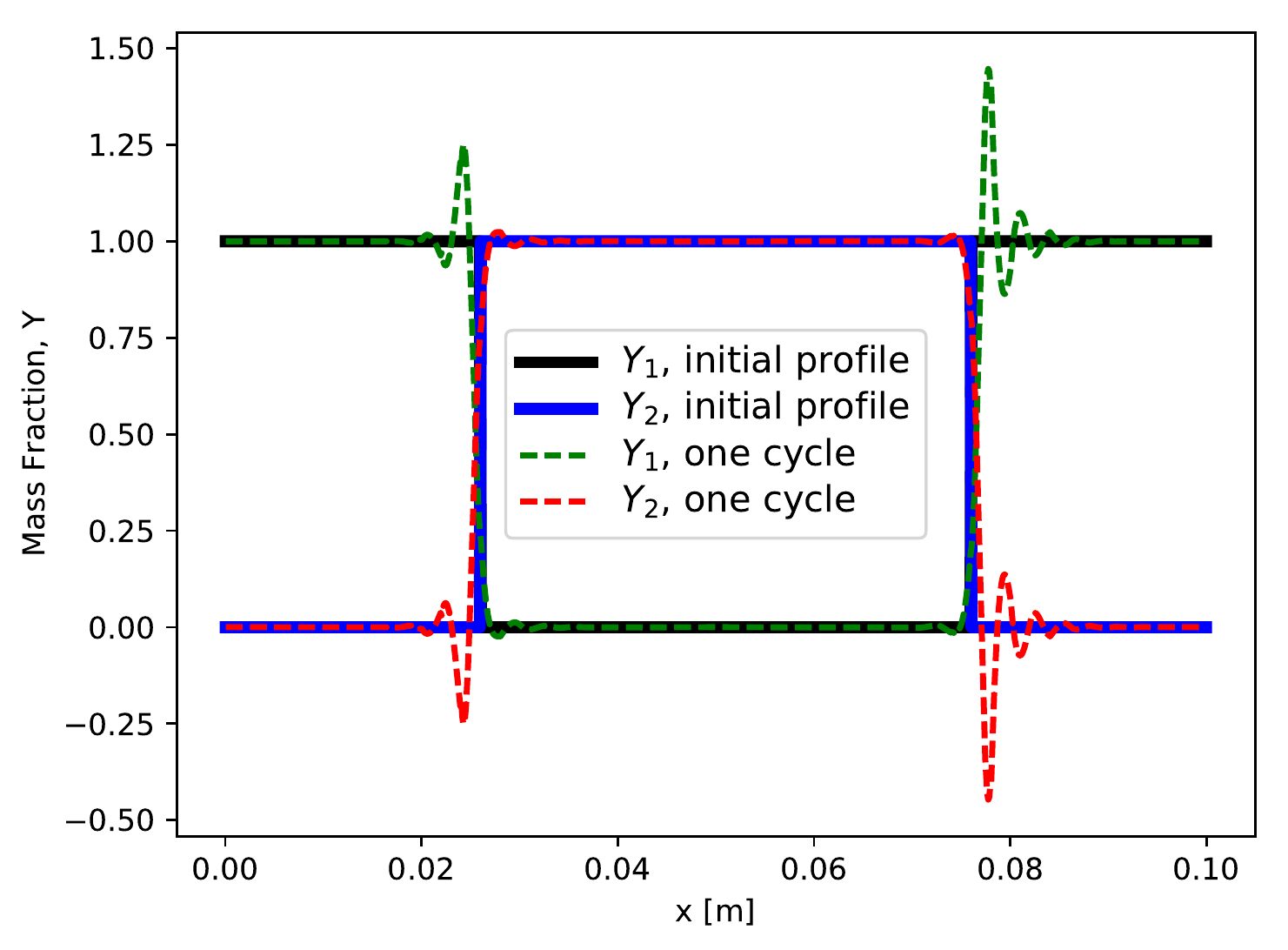}

}\caption{Mass fractions for species 1 and 2 for one cycle through domain using
the total energy formulation.\label{fig:species_discontinuity_constT-1}}
\end{figure}

A species discontinuity at constant temperature, pressure, and velocity,
is imposed with the following initial conditions

\begin{eqnarray*}
v & = & 10\textrm{ m/s},\\
Y_{1} & = & \begin{cases}
0 & \text{if }0.025<x<0.075\\
1 & \text{otherwise}
\end{cases},\\
Y_{2} & = & 1-Y_{1},\\
T & = & 300\textrm{ K},\\
p & = & 1\textrm{ atm}.
\end{eqnarray*}
Figs.~(\ref{fig:species_discontinuity_constT_nl_p1}), (\ref{fig:species_discontinuity_constT_nl_p2}),
and (\ref{fig:species_discontinuity_constT_nl_p3}) show the species
mass fractions for the $\mathrm{DG}(p=1)$, $\mathrm{DG}(p=2)$, and
$\mathrm{DG}(p=3)$ solutions using the total energy formulation after
one cycle. All three solutions present numerical overshoots and mixing
of the species mass fractions across the discontinuities. Some of
these numerical instabilities cause the mass fractions to be greater
than one or negative. As expected, the higher order solutions are
more oscillatory for the species mass fractions. Furthermore, larger
overshoots are present for the right hand side discontinuity as compared
to the left hand side discontinuity.

\begin{figure}[H]
\subfloat[Pressure \label{fig:species_discontinuity_constT_pressure}]{\includegraphics[width=0.45\columnwidth]{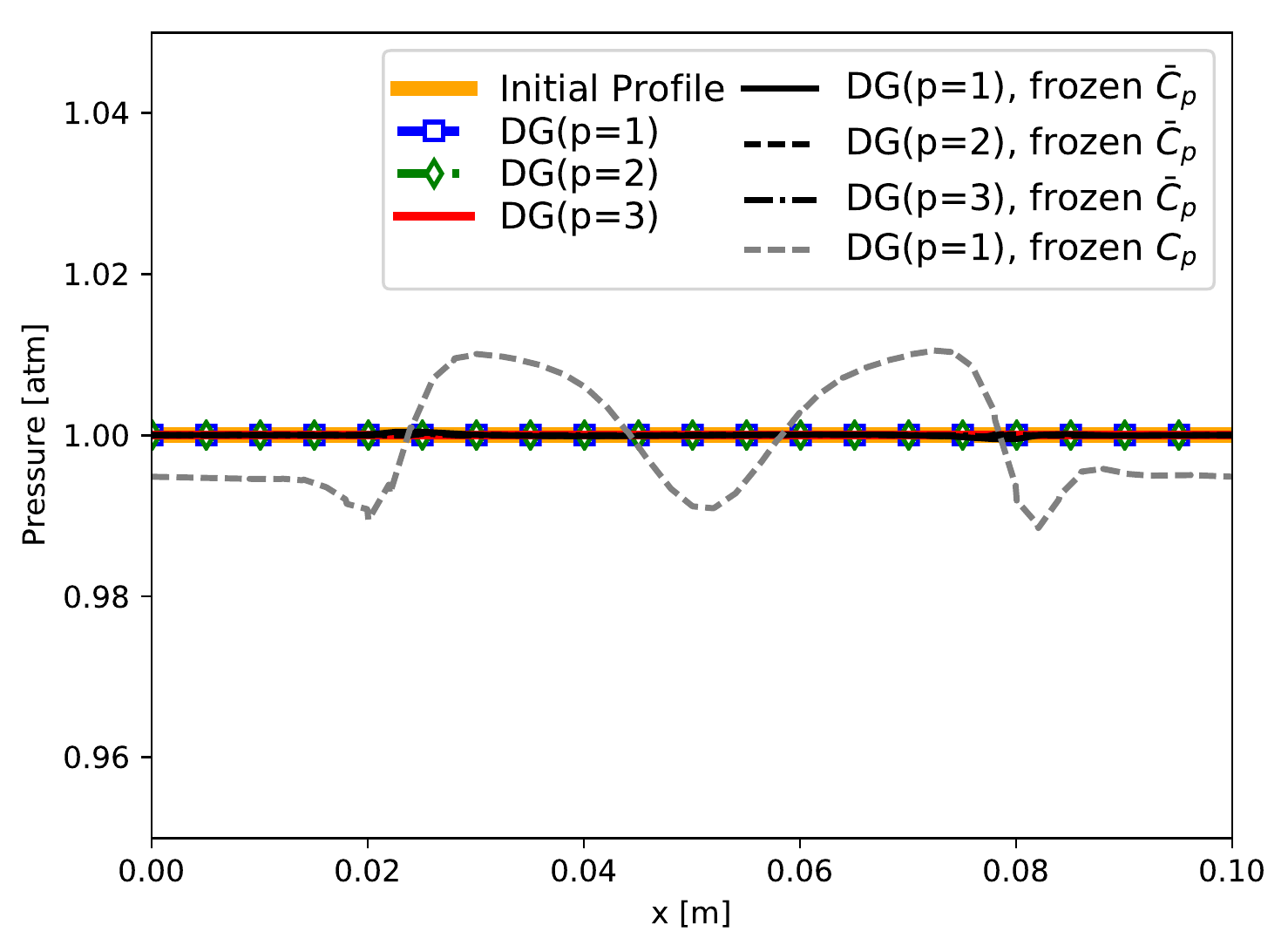}

}\hfill{}\subfloat[Pressure zoomed view \label{fig:species_discontinuity_constT_pressure_zoomed}]{\includegraphics[width=0.45\textwidth]{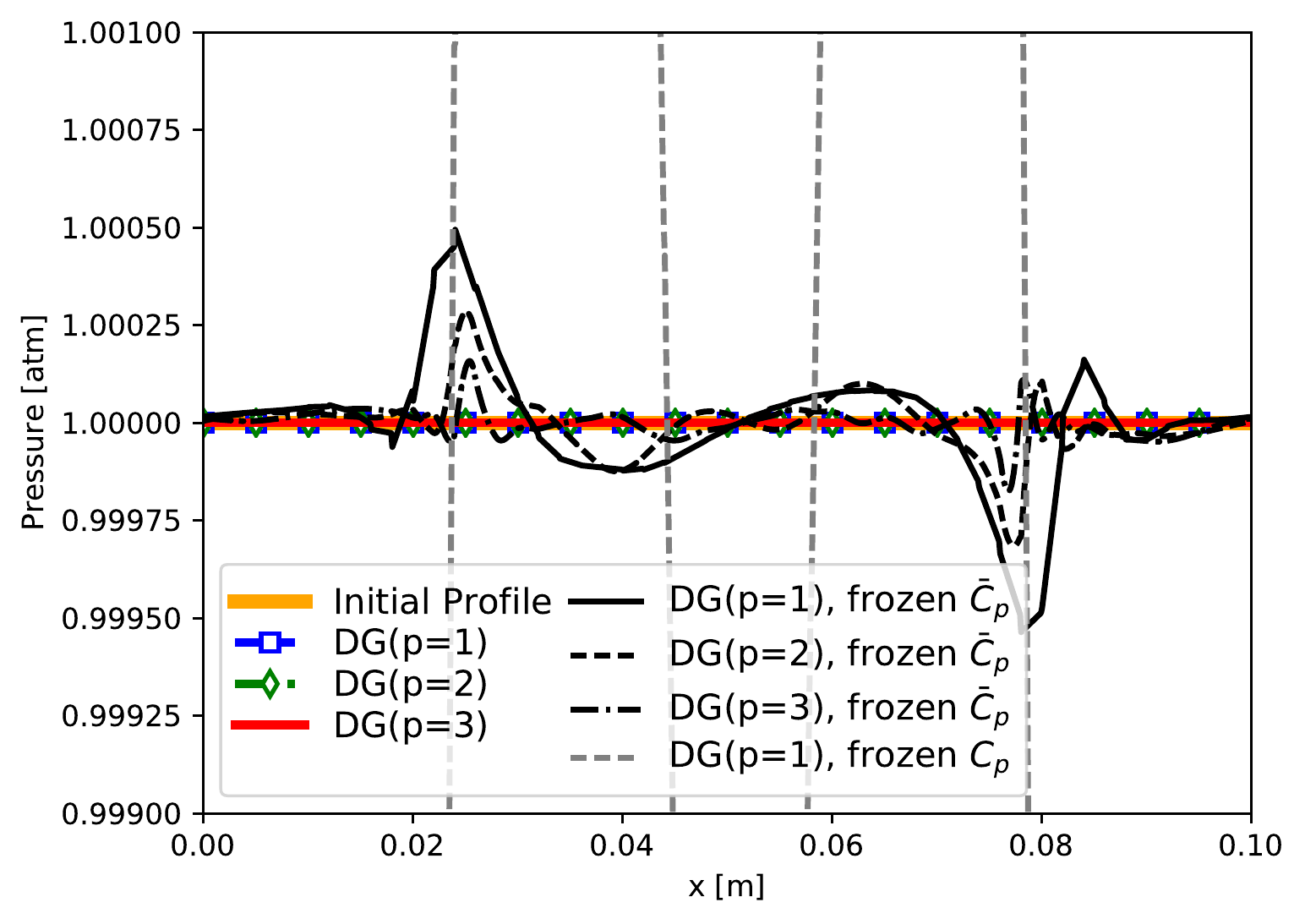}

}\caption{Pressure for $\mathrm{DG}(p=1)$, $\mathrm{DG}(p=2)$, and $\mathrm{DG}(p=3)$
solutions after one cycle. \label{fig:species_discontinuity_constT_p}}
\end{figure}

Fig.~(\ref{fig:species_discontinuity_constT_p}) shows the pressure
after one cycle for the total energy formulation and ratio of specific
heats formulations using frozen $\bar{C_{P}}$ and frozen $C_{p}$.
The frozen $C_{p}$ solution has pressure oscillations that are on
the order of three percent error. Small perturbations of pressure
caused by freezing $\bar{C_{p}}$ are shown in Fig.~(\ref{fig:species_discontinuity_constT_pressure_zoomed}).
The error caused by the frozen $\bar{C_{p}}$ formulation is reduced
as the approximation order is increased from $\mathrm{DG}(p=1)$ to
$\mathrm{DG}(p=3)$. In contrast, the pressure for each solution using
the total energy formulation maintains a flat profile, which is the
expected result from Section~\ref{sec:Discontinuities}. The pressure
in the total energy solutions never exceed an error of $\expnumber 1{-7}$
atm. 

\begin{figure}[H]
\subfloat[Temperature \label{fig:species_discontinuity_constT_temperature}]{\includegraphics[width=0.45\columnwidth]{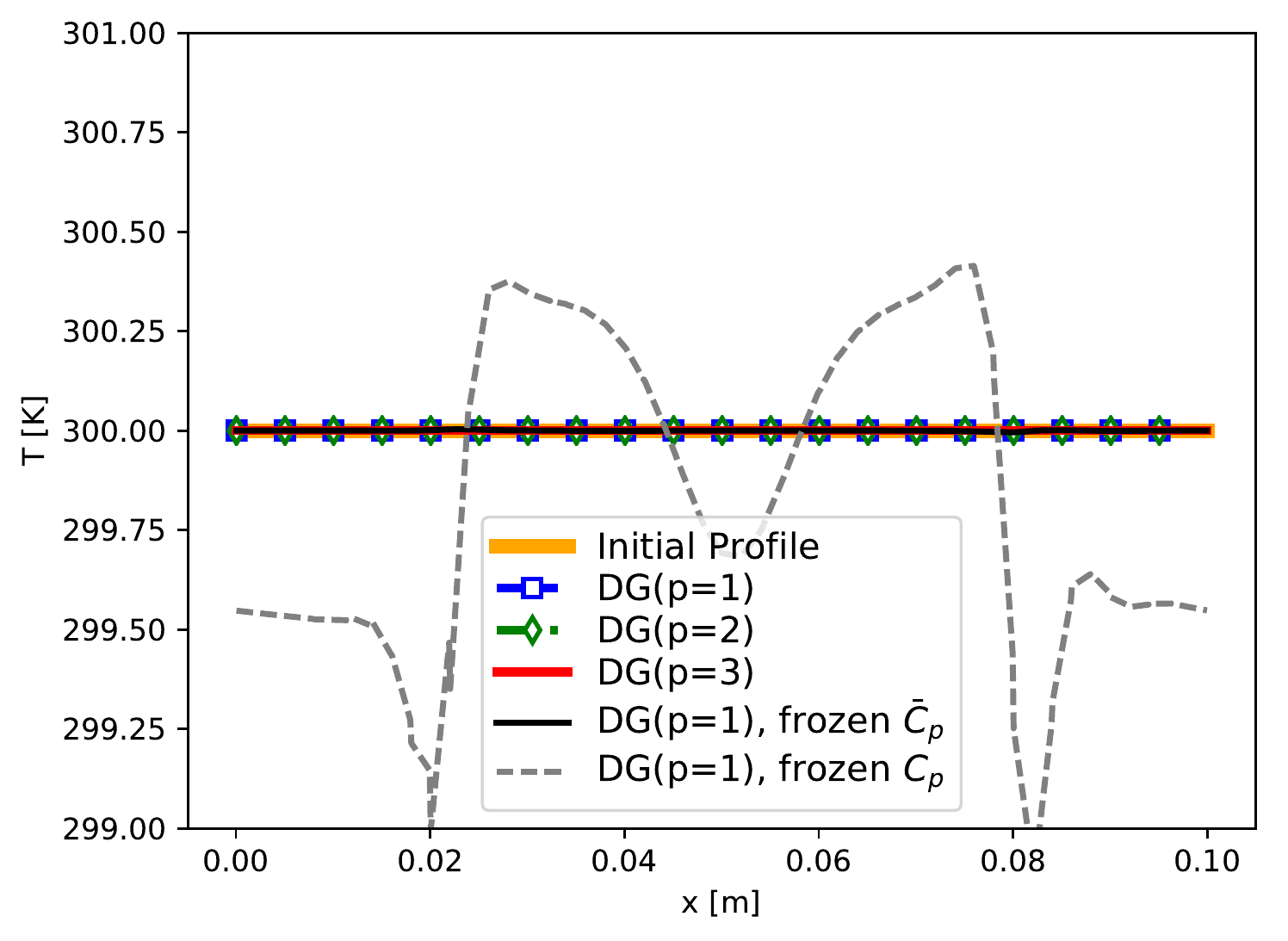}

}\hfill{}\subfloat[Temperature zoomed view \label{fig:species_discontinuity_constT_temperature_zoomed}]{\includegraphics[width=0.45\textwidth]{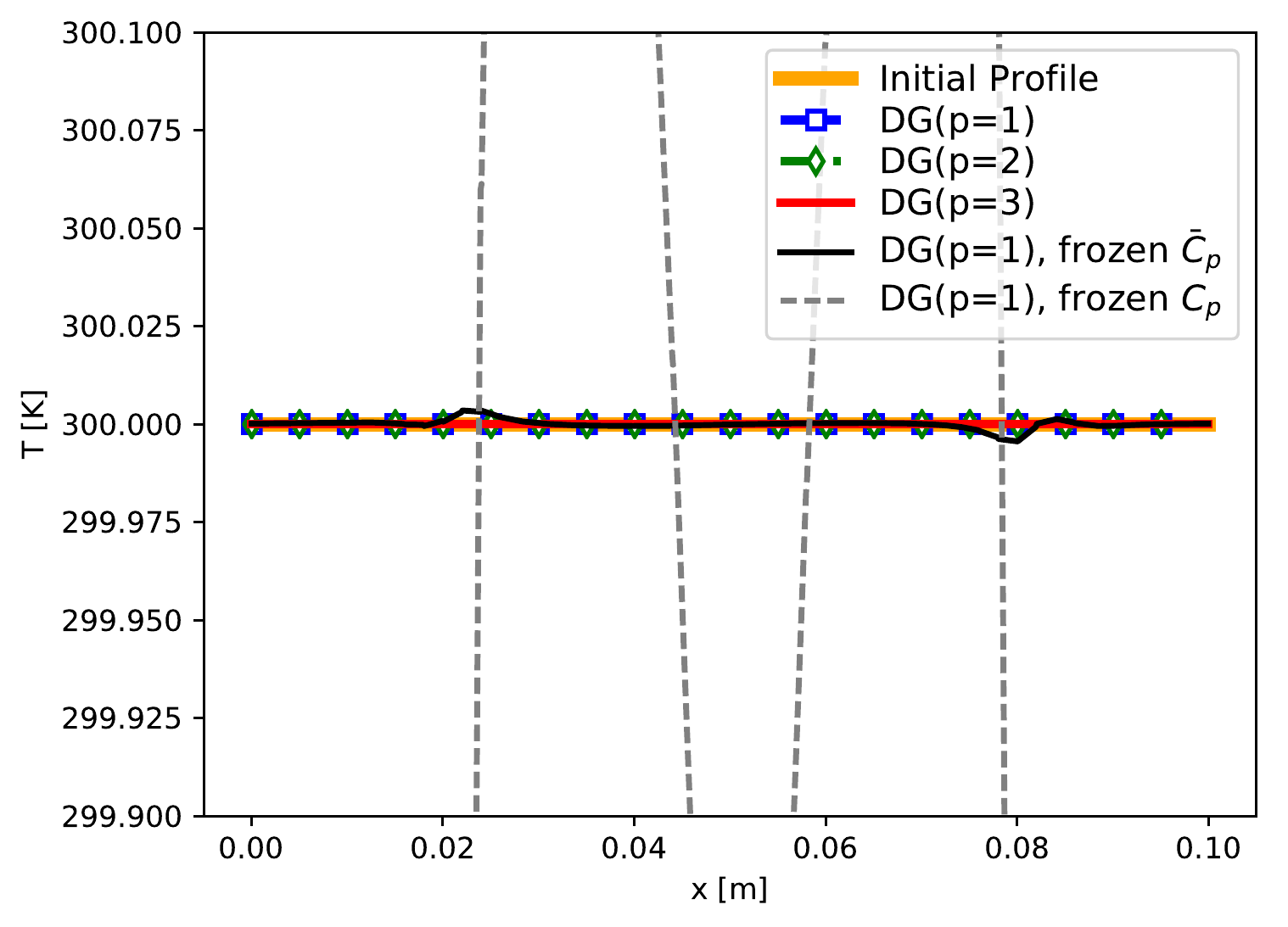}

}\caption{Temperature solutions after one cycle\label{fig:species_discontinuity_constT_T}}
\end{figure}

The corresponding temperature solutions are shown in Figs.~(\ref{fig:species_discontinuity_constT_temperature})~and~(\ref{fig:species_discontinuity_constT_temperature_zoomed}).
The frozen $C_{p}$ solution gives temperature fluctuations on the
order of $1$ K whereas the frozen $\bar{C_{p}}$ solution fluctuates
less than $0.025$ K. The total energy solution remains flat and does
not exceed $\expnumber 1{-4}$ K from the expected $300$ K.

\begin{figure}[H]
\subfloat[Pressure at $\mathrm{CFL}$ 0.1 and $\mathrm{CFL}$ 0.3 conditions
with frozen $\bar{C_{p}}$ formulation\label{fig:species_discontinuity_constT_pressure_cfl}]{\begin{centering}
\includegraphics[width=0.45\columnwidth]{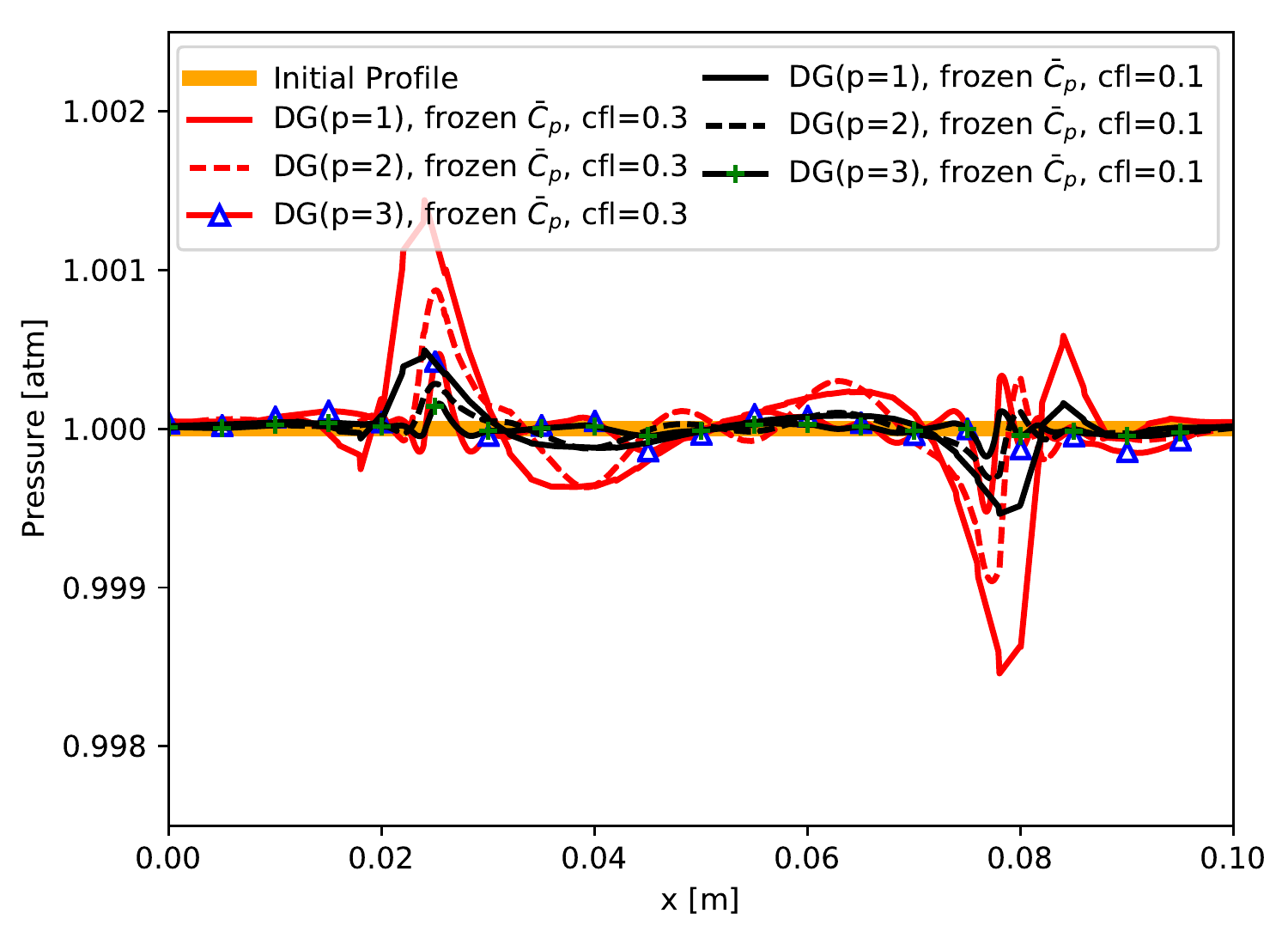}
\par\end{centering}
}\hfill{}\subfloat[Pressure after 100 cycles for $\mathrm{DG}(p=1)$ with total energy
and frozen $\bar{C_{p}}$ formulation\label{fig:species_discontinuity_constT_pressure_cycle}]{\begin{centering}
\includegraphics[width=0.45\columnwidth]{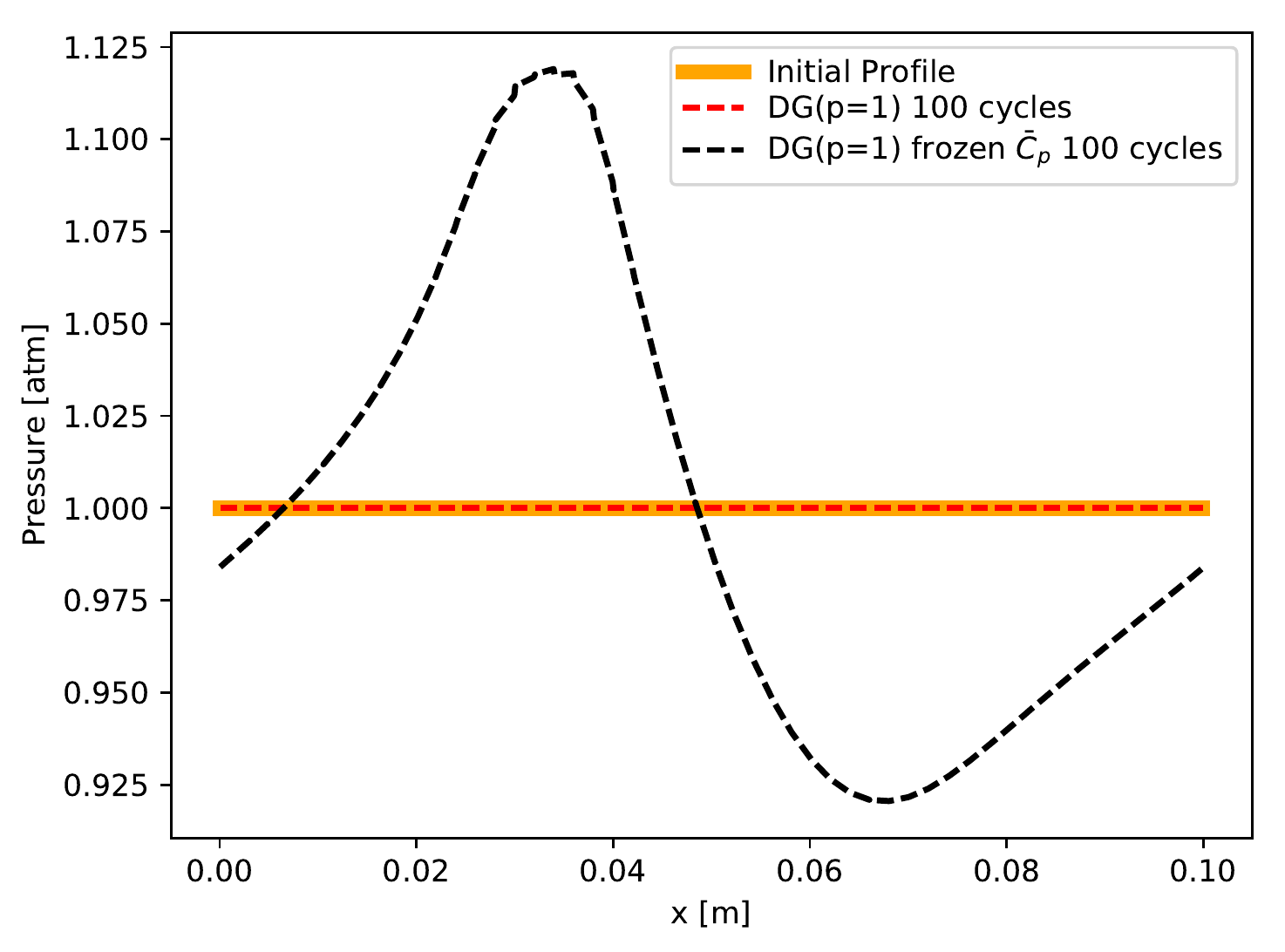}
\par\end{centering}
}
\centering{}\caption{Pressure for $\mathrm{DG}(p=1)$, $\mathrm{DG}(p=2)$, and $\mathrm{DG}(p=3)$
solutions after one cycle using $\mathrm{CFL}$ or 0.1 and $\mathrm{CFL}$
of 0.3 with frozen $\bar{C_{p}}$ and pressure after 100 cycles using
total energy formulation and frozen $\bar{C_{p}}$ for $\mathrm{DG}(p=1)$.\label{fig:species_discontinuity_constT_cfl_cycles}}
\end{figure}

Figure~(\ref{fig:species_discontinuity_constT_pressure_cfl}) shows
the solutions for $\mathrm{DG}(p=1)$, $\mathrm{DG}(p=2)$, and $\mathrm{DG}(p=3)$
for frozen $\bar{C_{p}}$ at $\mathrm{CFL}$ of $0.1$ and $0.3$.
The larger time steps exacerbate the instability introduced by freezing
$\bar{C_{p}}$ to exceed 0.001 atm which is still an order of magnitude
less than the error of the frozen $C_{p}$ solutions. The solutions
for the total energy formulation is unaffected by the time step and
therefore the corresponding results are not shown.

Figure~(\ref{fig:species_discontinuity_constT_pressure_cycle}) shows
the pressure solution of the $\mathrm{DG}(p=1)$ solution for the
frozen $\bar{C_{p}}$ formulation and total energy formulation after
100 cycles, i.e. $t=1$ s. The pressure oscillations for the 100 cycle
solution using the frozen $\bar{C_{p}}$ formulation grow in time
to be on the order of the one cycle frozen $C_{p}$ formulation, whereas
the total energy formulation remains constant after 100 cycles.

\subsubsection{Species Discontinuities and a Continuous Temperature Variation\label{subsec:species-discontinuity-temperature-gradient}}

\begin{figure}[H]
\subfloat[$\mathrm{DG}\left(p=1\right)$ solution \label{fig:species_discontinuity_gradT_p1}]{\includegraphics[width=0.32\columnwidth]{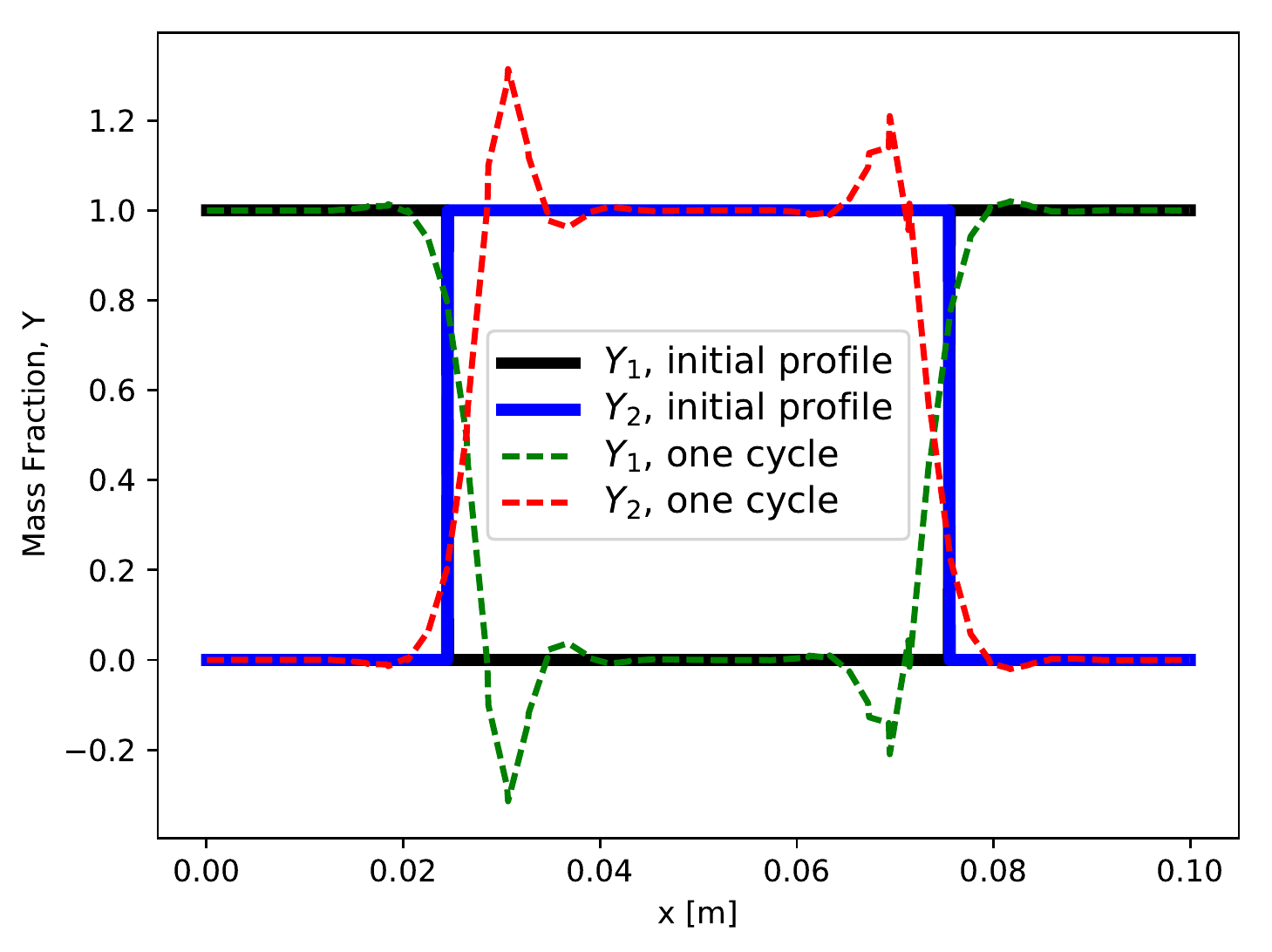}

}\hfill{}\subfloat[$\mathrm{DG}\left(p=2\right)$ solution \label{fig:species_discontinuity_gradT_p2}]{\includegraphics[width=0.32\textwidth]{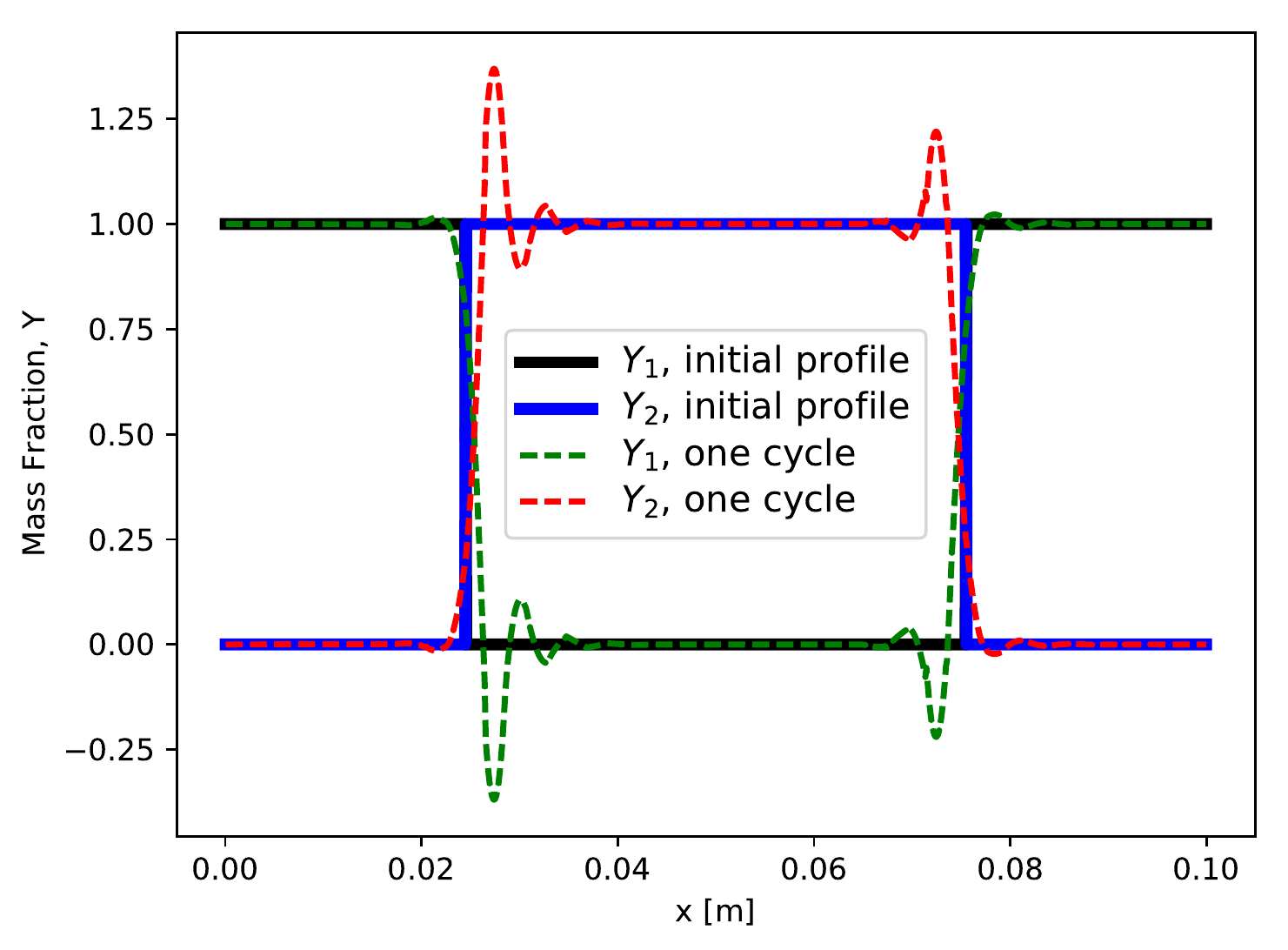}

}\hfill{}\subfloat[$\mathrm{DG}\left(p=3\right)$ solution \label{fig:species_discontinuity_gradT_p3}]{\includegraphics[width=0.32\textwidth]{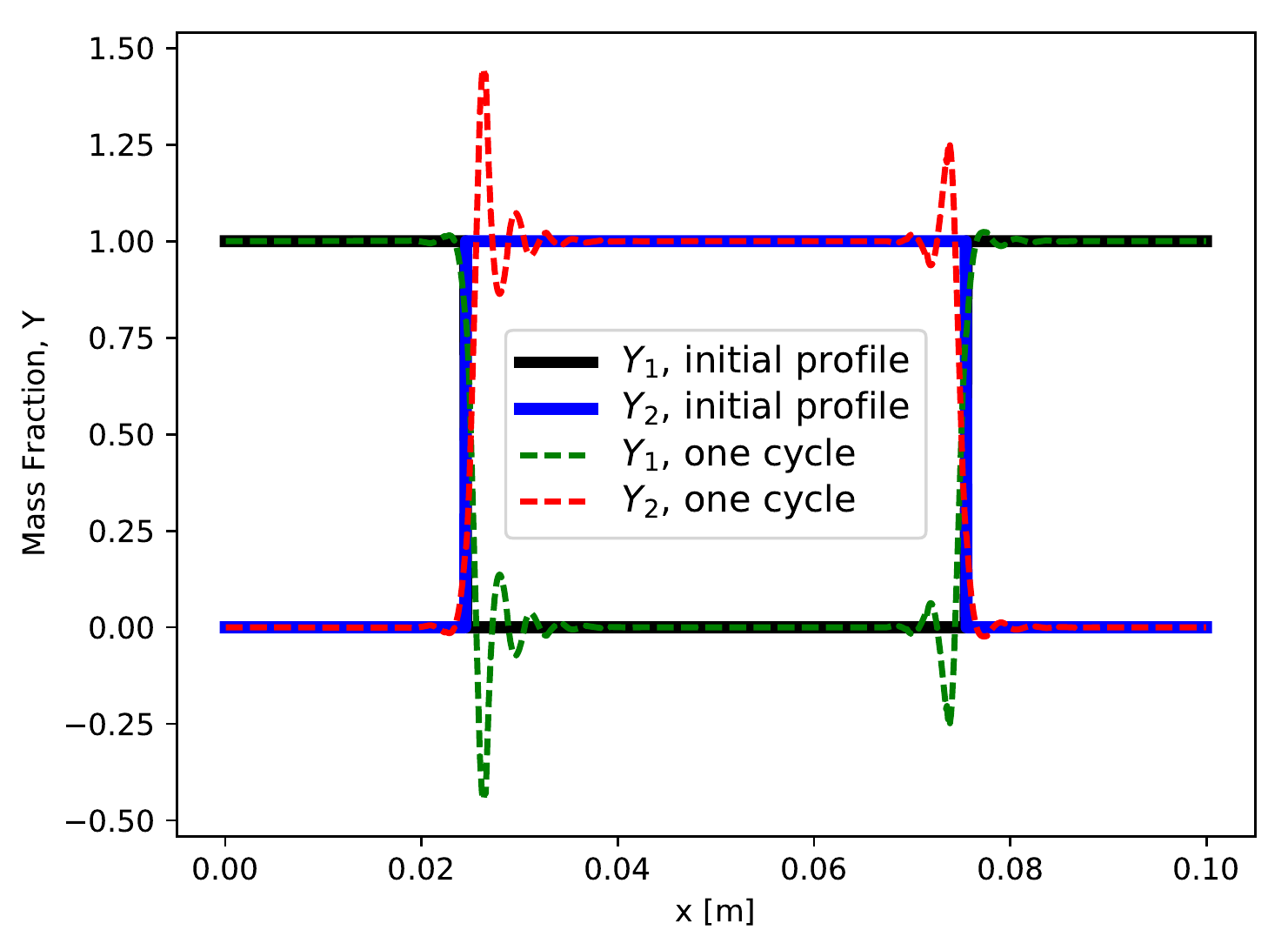}

}\caption{Mass fractions for species 1 and 2 for one cycle through domain using
the total energy formulation.\label{fig:species_discontinuity_gradT}}
\end{figure}

A sinusoidal variation in temperature is introduced to the initial
profile described in Section~(\ref{subsec:species-discontinuity-constant-temperature}).
The initial conditions are given as follows

\begin{eqnarray*}
v & = & 10\textrm{ m/s},\\
Y_{1} & = & \begin{cases}
0 & \text{if }0.025<x<0.075\\
1 & \text{otherwise}
\end{cases},\\
Y_{2} & = & 1-Y_{1},\\
T & = & 350+50\sin\left(20\pi x\right)\textrm{ K},\\
p & = & 1\textrm{ atm}.
\end{eqnarray*}
Figs.~(\ref{fig:species_discontinuity_gradT_p1}), (\ref{fig:species_discontinuity_gradT_p2}),
and (\ref{fig:species_discontinuity_gradT_p3}) show the species mass
fractions for $\mathrm{DG}(p=1)$, $\mathrm{DG}(p=2)$, and $\mathrm{DG}(p=3)$
solutions after one cycle. Similar to the previous test case, the
three solutions present numerical overshoots and mixing of the species
mass fractions across the discontinuities. Here the left side discontinuity
has larger overshoots in the higher temperature region as compared
to the right hand side discontinuity. 

\begin{figure}[H]
\subfloat[Pressure \label{fig:species_discontinuity_gradT_pressure}]{\includegraphics[width=0.45\columnwidth]{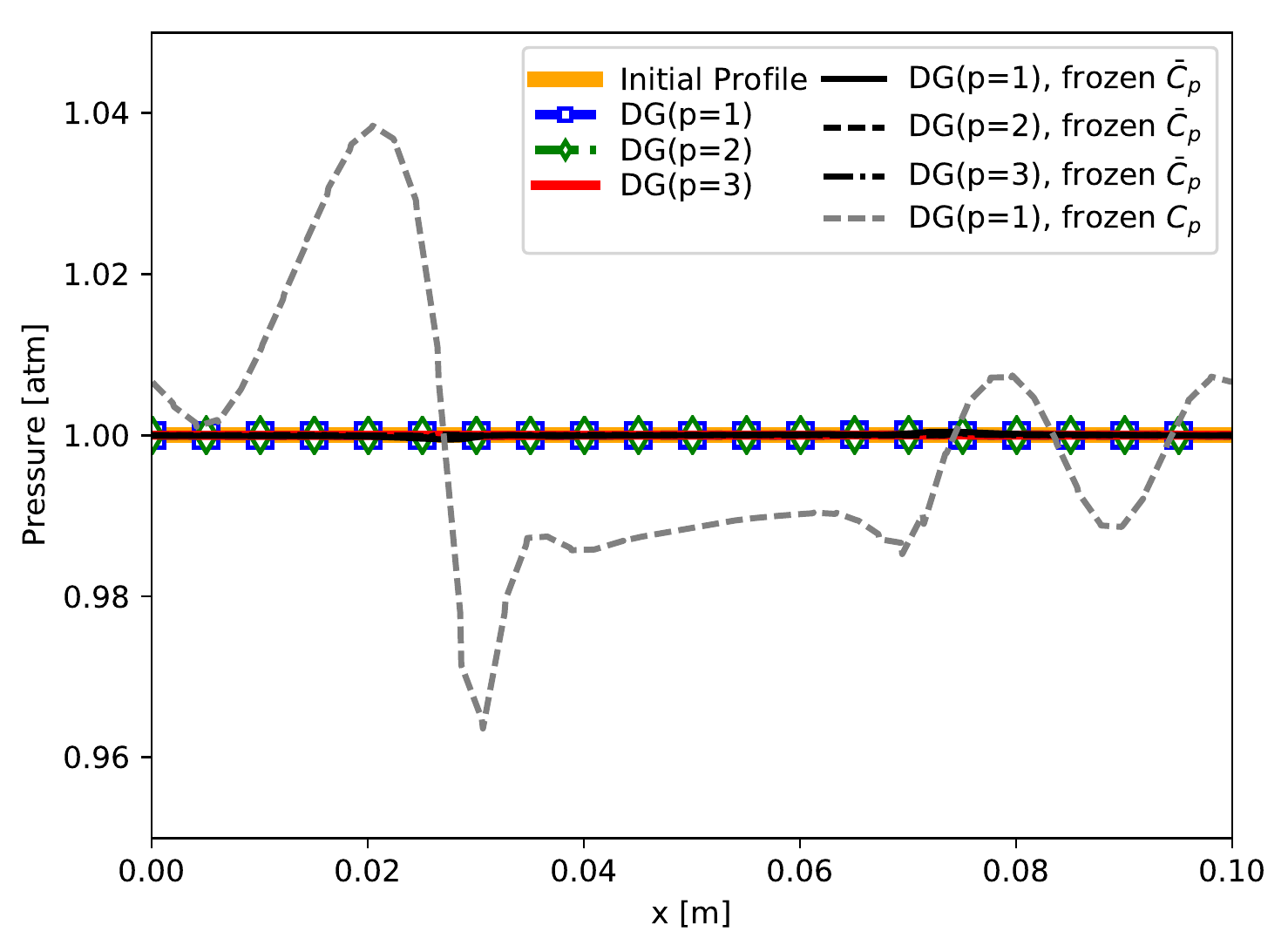}

}\hfill{}\subfloat[Pressure zoomed view \label{fig:species_discontinuity_gradT_pressure_zoomed}]{\includegraphics[width=0.45\textwidth]{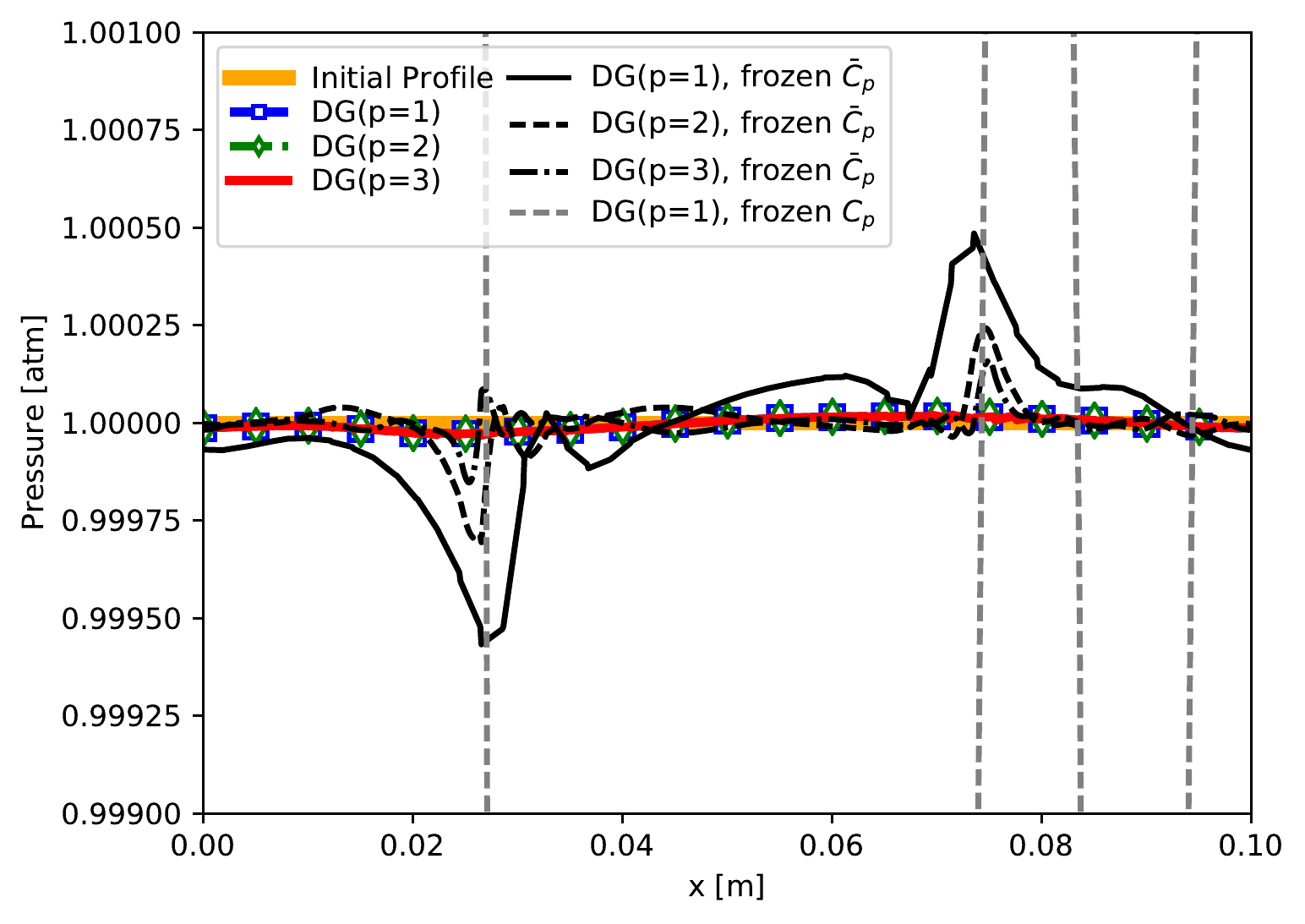}

}\caption{Pressure for $\mathrm{DG}(p=1)$, $\mathrm{DG}(p=2)$, and $\mathrm{DG}(p=3)$
solutions after one cycle. \label{fig:species_discontinuity_gradT_p}}
\end{figure}

Figure~(\ref{fig:species_discontinuity_gradT_p}) shows the pressure
for all three solutions using the total energy formulation and ratio
of specific heats formulation with frozen $C_{p}$ and frozen $\bar{C_{p}}$
after one cycle. Fig.~(\ref{fig:species_discontinuity_gradT_pressure})
shows the large oscillations like the previous test case for the frozen
$C_{p}$ formulation and smaller oscillations using the frozen $\bar{C_{p}}$
formulation. The oscillations caused by freezing $\bar{C_{p}}$ are
reduced as the accuracy of the approximation is increased from $\mathrm{DG}(p=1)$
to $\mathrm{DG}(p=3)$. Again, the pressure for each solution using
the total energy formulation maintains a flat pressure profile, even
with a spatially varying temperature profile. This is expected as
the temperature is continuous through the domain. 

\begin{figure}[H]
\centering{}\includegraphics[width=0.45\columnwidth]{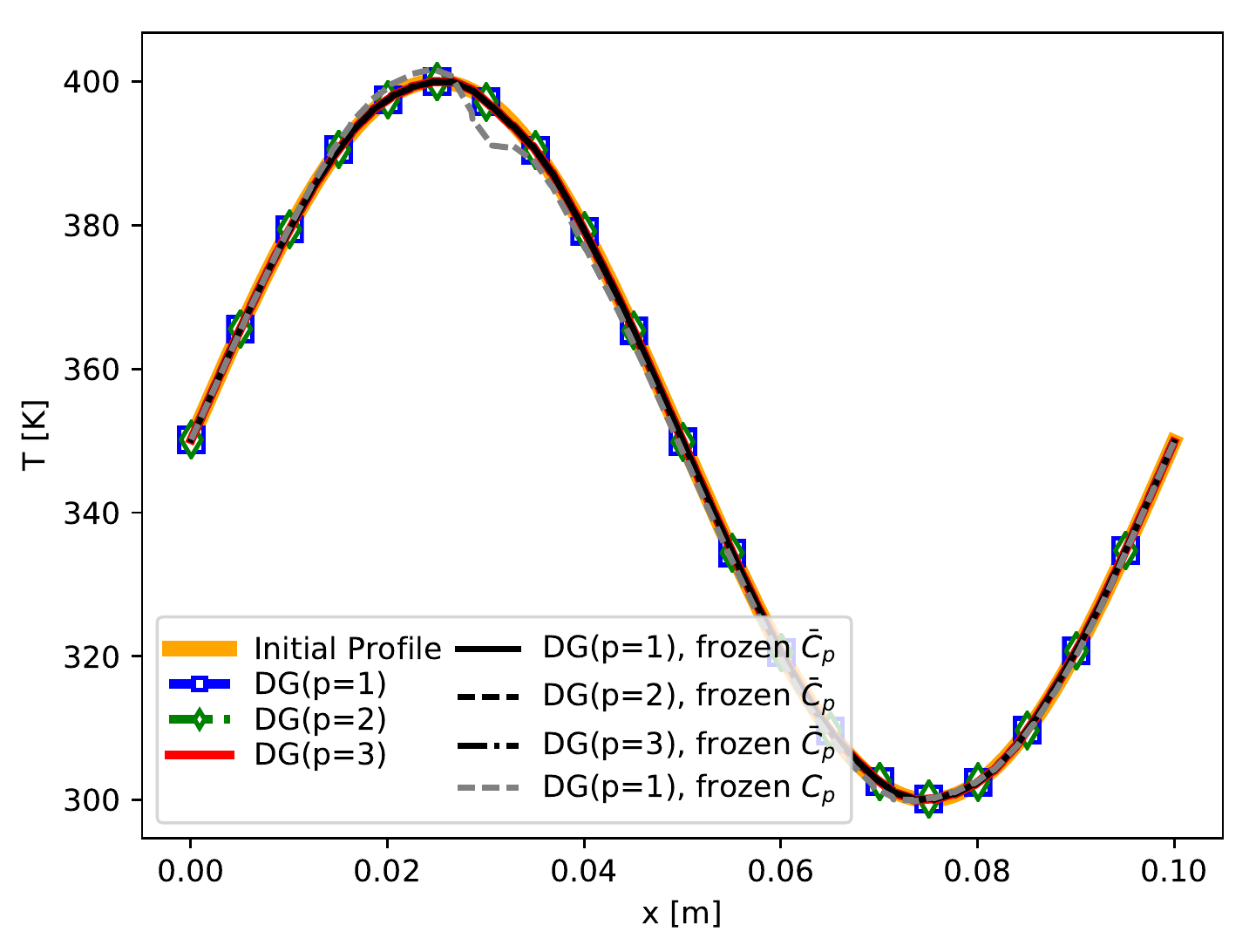}\caption{Temperature for $\mathrm{DG}(p=1)$, $\mathrm{DG}(p=2)$, and $\mathrm{DG}(p=3)$
solutions after one cycle.\label{fig:species_discontinuity_gradT_T}}
\end{figure}

Figure~(\ref{fig:species_discontinuity_gradT_T}) shows the computed
temperature from solutions using the total energy formulation and
the ratio of specific heats formulation with frozen $C_{p}$ and frozen
$\bar{C_{p}}$ after one cycle. For the total energy and ratio of
specific heats formulation with frozen $\bar{C}_{p}$ the temperature
is within $\expnumber 1{-4}$ K of the analytical sinusoidal solution.
The ratio of specific heats formulation with frozen $C_{p}$ is also
shown and departs from the analytical result with the largest deviation
of $5$ K in the higher temperature region


\subsubsection{Species Discontinuities with Temperature Discontinuities\label{subsec:species-discontinuity-temperature-discontinuity}}

\begin{figure}[H]
\subfloat[$\mathrm{DG}\left(p=1\right)$ solution \label{fig:species_discontinuity_discT_p1}]{\includegraphics[width=0.32\columnwidth]{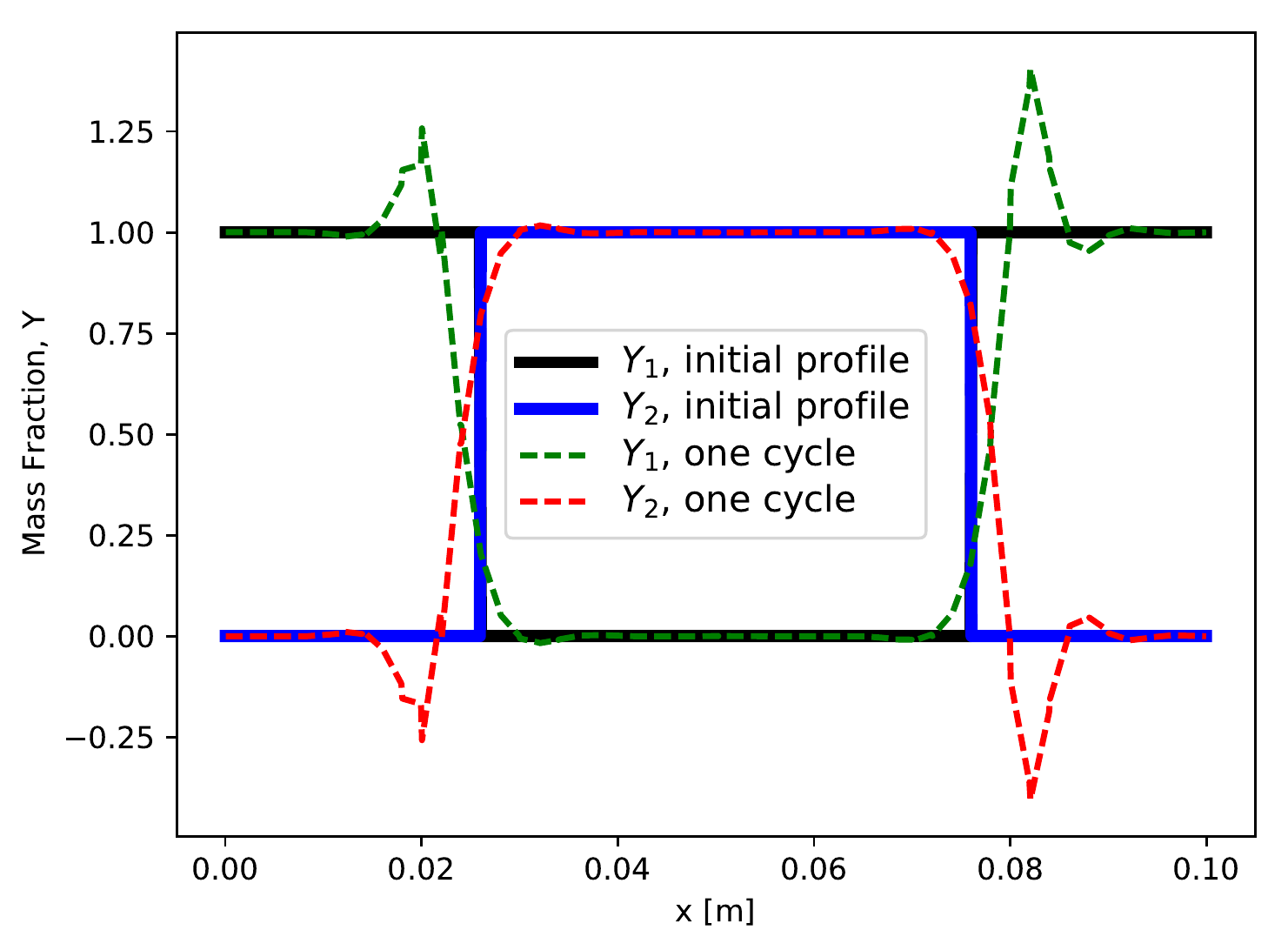}

}\hfill{}\subfloat[$\mathrm{DG}\left(p=2\right)$ solution \label{fig:species_discontinuity_discT_p2}]{\includegraphics[width=0.32\textwidth]{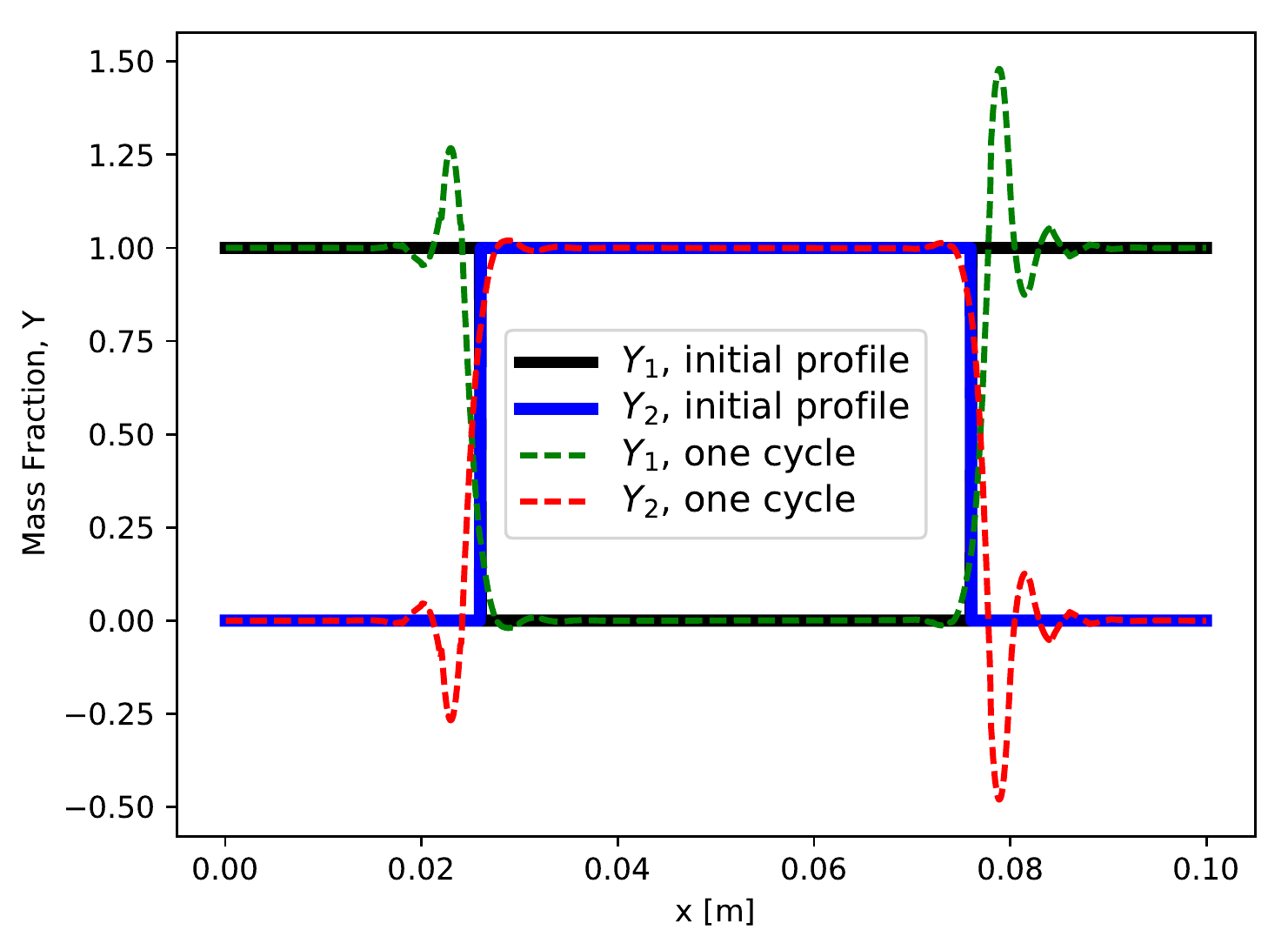}

}\hfill{}\subfloat[$\mathrm{DG}\left(p=3\right)$ solution \label{fig:species_discontinuity_discT_p3}]{\includegraphics[width=0.32\textwidth]{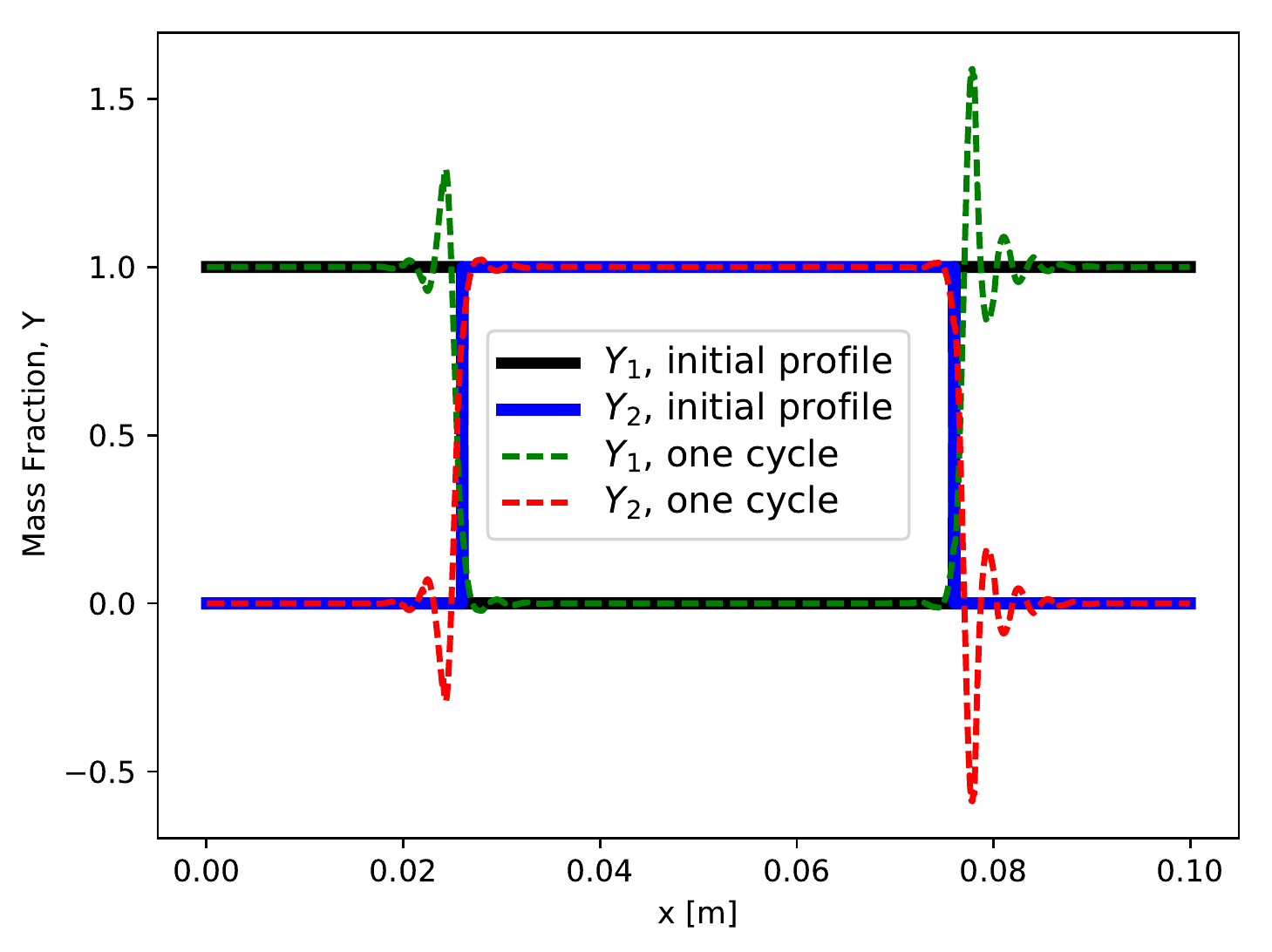}

}\caption{Mass fractions for species 1 and 2 for one cycle through domain using
the total energy formulation.\label{fig:species_discontinuity_discT}}
\end{figure}

\begin{figure}[H]
\subfloat[Pressure solutions after one cycle. \label{fig:species_discontinuity_discT_p}]{\includegraphics[width=0.45\columnwidth]{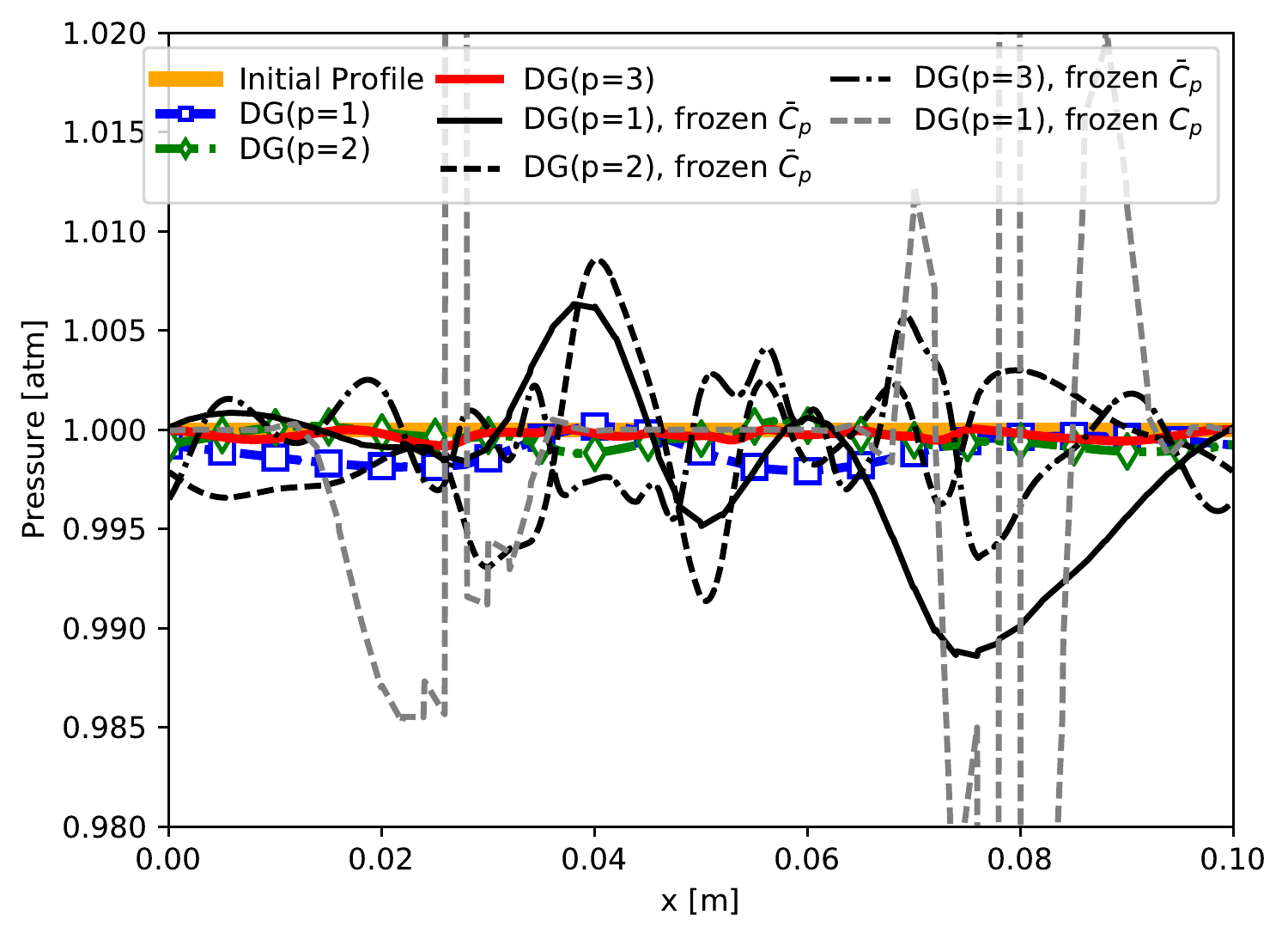}

}\hfill{}\subfloat[Temperature solutions after one cycle.\label{fig:species_discontinuity_discT_T}]{\includegraphics[width=0.45\textwidth]{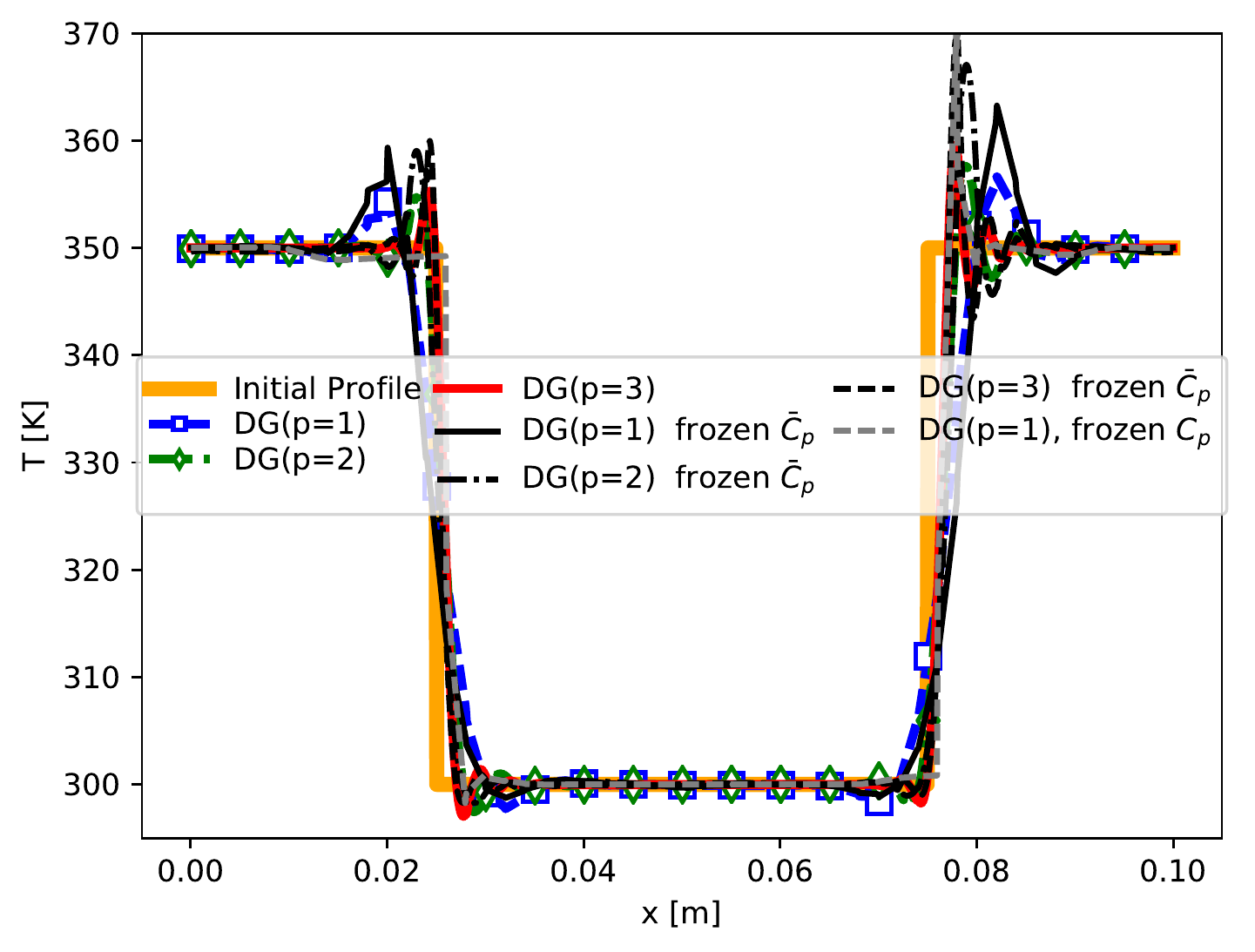}

}\caption{Pressure and temperature for $\mathrm{DG}(p=1)$, $\mathrm{DG}(p=2)$,
and $\mathrm{DG}(p=3)$ solutions after one cycle. The frozen $C_{p}$
solution at 100 steps is shown in grey.\label{fig:species_discontinuity_discT_p_T}}
\end{figure}

The following test case uses the initial conditions described in Section~\ref{subsec:species-discontinuity-constant-temperature},
except the initial temperature profile is piecewise constant instead
of constant. The initial conditions are given as follows

\begin{eqnarray*}
v & = & 10\textrm{ m/s},\\
Y_{1} & = & \begin{cases}
0 & \text{if }0.025<x<0.075\\
1 & \text{otherwise}
\end{cases},\\
Y_{2} & = & 1-Y_{1},\\
T & = & \begin{cases}
300\text{ K} & \text{if }0.025<x<0.075\\
350\text{ K} & \text{otherwise}
\end{cases},\\
p & = & 1\textrm{ atm}.
\end{eqnarray*}
Figs.~\ref{fig:species_discontinuity_discT_p1}, \ref{fig:species_discontinuity_discT_p2},
and \ref{fig:species_discontinuity_discT_p3} show the species mass
fractions for $\mathrm{DG}(p=1)$, $\mathrm{DG}(p=2)$, and $\mathrm{DG}(p=3)$
solutions after one cycle. Similar to the previous test case, the
three solutions present numerical overshoots and mixing of the species
mass fractions across the discontinuities. Figs.~\ref{fig:species_discontinuity_discT_p}
and \ref{fig:species_discontinuity_discT_T} show the pressure and
temperature, respectively, after one cycle for the total energy formulation
and the ratio of specific heats formulation with frozen $\bar{C_{P}}$.
The ratio of specific heats formulation with frozen $C_{p}$ simulation
fails before one complete cycle. The solution corresponding to frozen
$C_{p}$ is shown after 100 time steps. The pressure for each solution
using the total energy formulation causes pressure oscillations that
are an order of magnitude less than the frozen $\bar{C_{p}}$ simulations.
The oscillations in the total energy formulation are expected based
on the discussion in Section~\ref{sec:Discontinuities}. Both the
total energy formulation and the ratio of specific heats formulation
with frozen $\bar{C_{p}}$ produce overshoots and undershoots at the
temperature discontinuities. The temperature oscillations associated
with the ratio of specific heats formulation with frozen $\bar{C_{p}}$
are larger than the oscillations corresponding to the total energy
formulation.

\begin{figure}[H]
\subfloat[Temperature solutions after 100 cycles.\label{fig:species_discontinuity_discT_100_T}]{\includegraphics[width=0.45\textwidth]{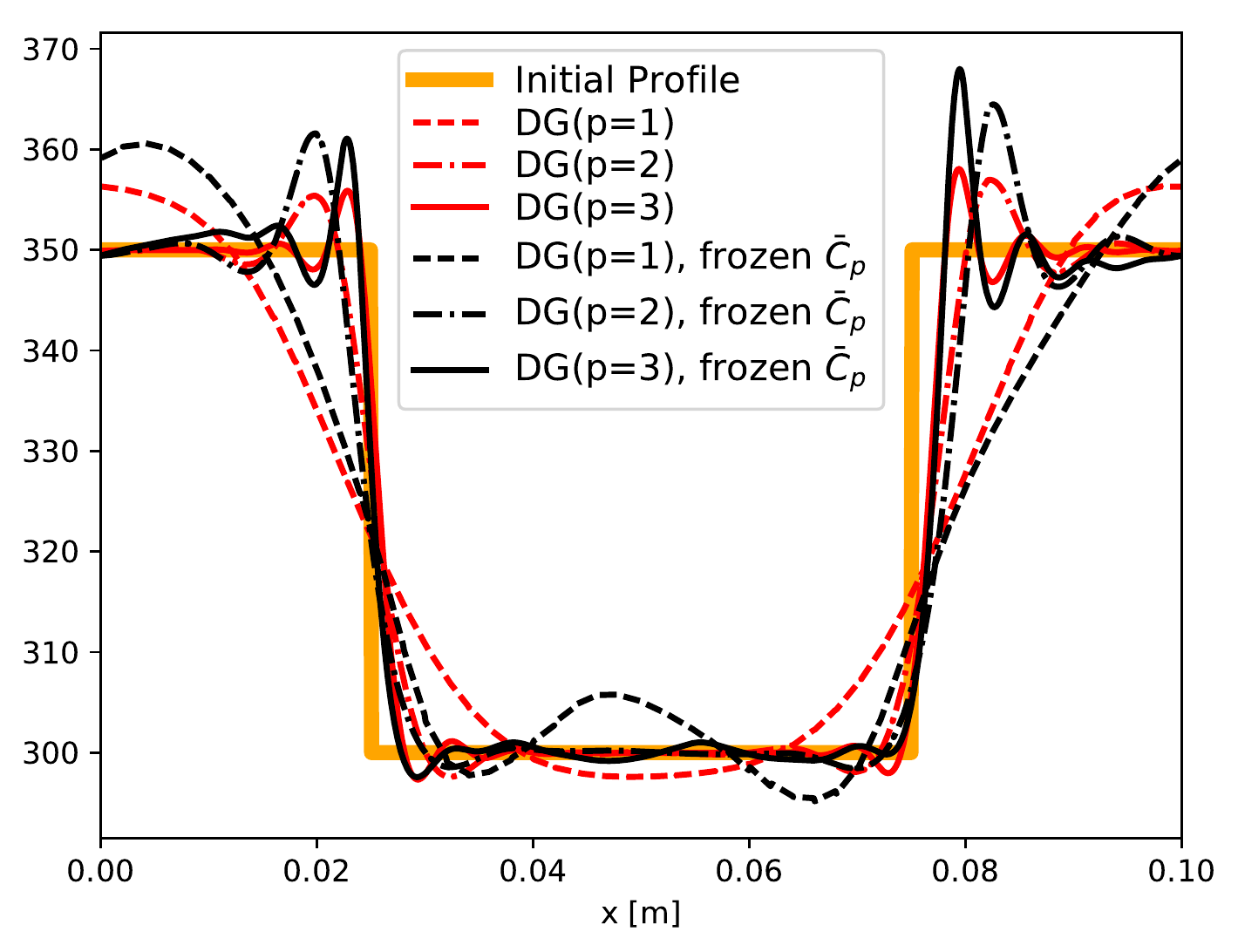}

}\hfill{}\subfloat[Pressure solutions after 100 cycles. \label{fig:species_discontinuity_discT_100_p}]{\includegraphics[width=0.45\columnwidth]{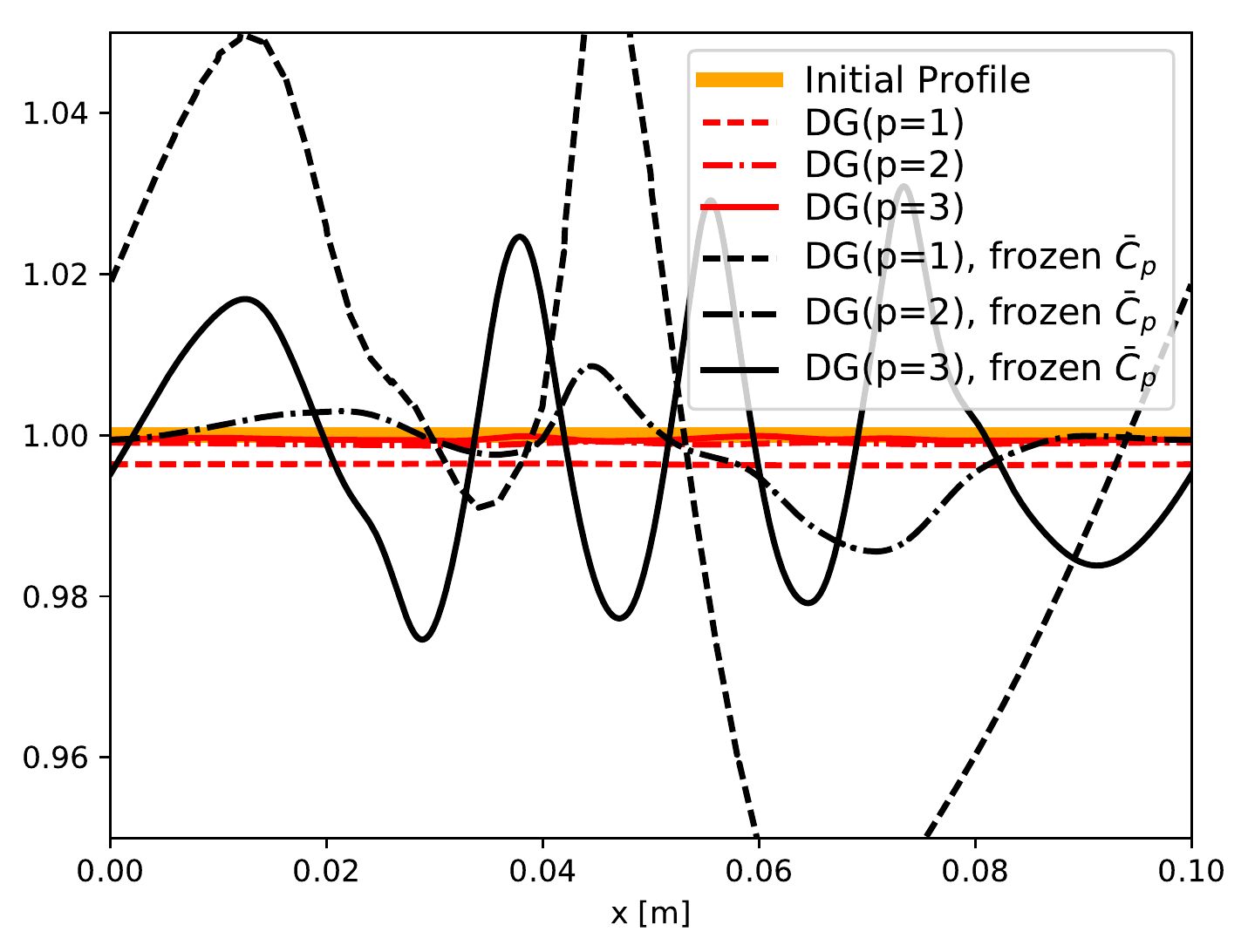}

}\caption{Pressure and temperature for $\mathrm{DG}(p=1)$, $\mathrm{DG}(p=2)$,
and $\mathrm{DG}(p=3)$ solutions after 100 cycles. }
\end{figure}

Figs.~\ref{fig:species_discontinuity_discT_100_T} and \ref{fig:species_discontinuity_discT_100_p}
show the temperature and pressure solutions, respectively, for the
total energy formulation and the ratio of specific heats formulation
with frozen $\bar{C_{P}}$ after 100 cycles, i.e. $t=1$ s. The temperature
profiles become more diffuse with the larger number of cycles (see
Fig.~\ref{fig:species_discontinuity_discT_T} for the comparison
of 1 cycle). The pressure oscillations for the total energy formulation
cause the ambient pressure to fall below 1 atm. This departure from
ambient was improved by increasing the approximation order from $\mathrm{DG}(p=1)$
to $\mathrm{DG}(p=3)$. The pressure oscillations for the 100 cycle
solution using the frozen $\bar{C_{p}}$ formulation grow in time
regardless of approximation order (see Fig.~\ref{fig:species_discontinuity_discT_p}
for the one cycle solution for frozen $\bar{C_{P}}$). Furthermore,
the frozen $\bar{C}_{p}$ formulation oscillations are an order of
magnitude larger than the oscillations present in the total energy
formulation.

\subsection{Continuous Simulations }

\subsubsection{Thermal Bubble}

\begin{figure}[H]
\subfloat[Pressure solution after one cycle. \label{fig:tb_1D_lcp_p}]{\includegraphics[width=0.45\columnwidth]{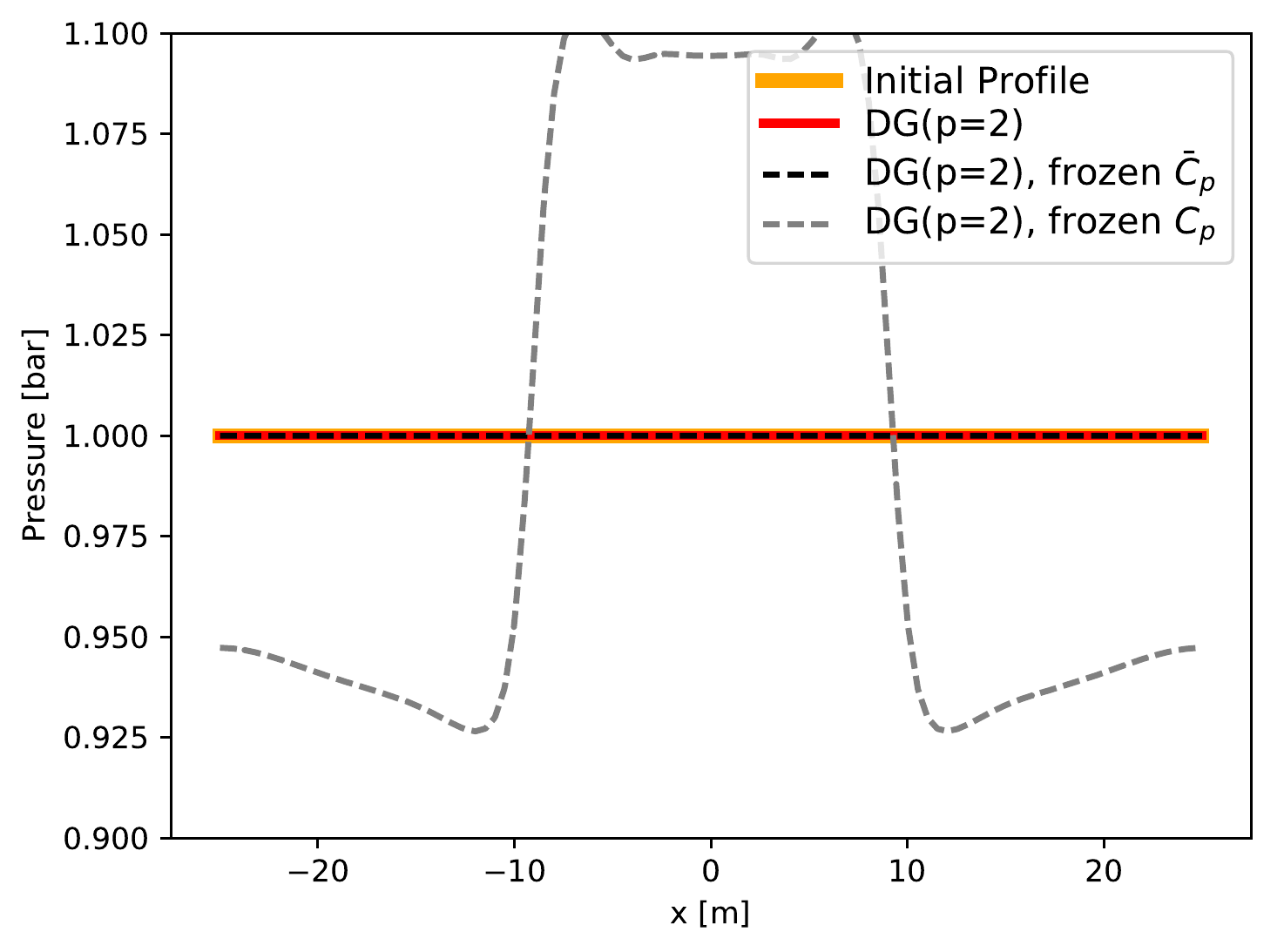}

}\hfill{}\subfloat[Temperature solution after one cycle. \label{fig:tb_1D_lcp_T}]{\includegraphics[width=0.45\columnwidth]{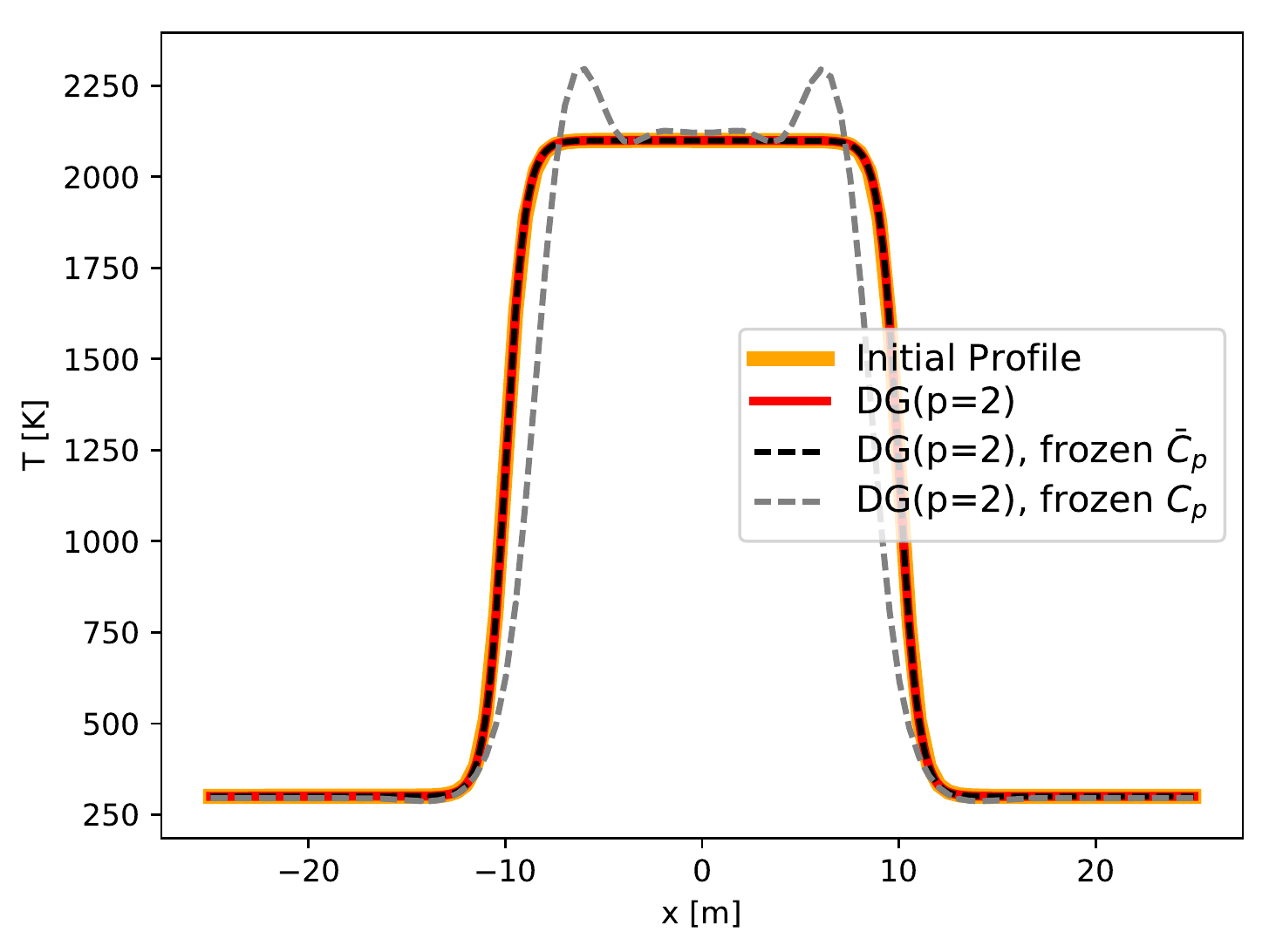}

}%

\caption{Thermal bubble simulation results corresponding to $\mathrm{DG}\left(p=2\right)$
after one cycle.\label{fig:tb_1D_lcp}}
\end{figure}

Here we present the one dimensional thermal bubble test case previously
presented by~\citep{Lv15}. For this test case, a periodic domain
$50$ m domain, $(-25,25)$ m, with grid spacing, $h$, of 0.5 m is
used with the following initial conditions

\begin{eqnarray*}
v & = & 1\textrm{ m/s},\\
Y_{H2} & = & \frac{1}{2}\left[1-\tanh\left(|x|-10\right)\right],\\
Y_{O_{2}} & = & 1-Y_{H_{2}},\\
T & = & 1200-900\tanh\left(|x|-10\right)\textrm{ K},\\
p & = & 1\textrm{ bar}.
\end{eqnarray*}
The test case is run for 1 cycle, $t=50$ s, using $\mathrm{DG}(p=2)$
with $\mathrm{CFL}=0.1$ and the inviscid non-reacting formulation
of Eqs.~(\ref{eq:conservation-law-strong-form})- (\ref{eq:conservation-law-boundary-condition}).
No artificial viscosity or fail-safe limiting is used in this test
case. The mesh resolution was too coarse to stably compute a $\mathrm{DG}(p=1)$
solution without limiting. Like the previous test cases, the analytical
solution after one cycle is the same as the same as the initial profile.
Figs.~\ref{fig:tb_1D_lcp_p} and \ref{fig:tb_1D_lcp_T} show the
results for pressure and temperature, respectively. The pressure is
constant throughout for both the total energy formulation and the
ratio of specific heats formulation with frozen $\bar{C_{p}}$ with
variations on the order of $\expnumber 1{-5}\,$ atm. The pressure
for the ratio of specific heats formulation with frozen $C_{p}$ fluctuates
on the order of $10$\% of the expected ambient pressure. Previous
work reported that without the double flux method the pressure fluctuated
throughout the solution by $3\%$ of the expected ambient pressure~\citep{Lv15}.
These issues do not occur in the solutions corresponding to the total
energy formulation.

\subsubsection{One-Dimensional Premixed Flame}

We use the total energy formulation with detailed chemical kinetics
and transport in the viscous, reacting formulation of Eqs.~(\ref{eq:conservation-law-strong-form})-
(\ref{eq:conservation-law-boundary-condition}) to approximate the
solution to a one-dimensional premixed flame using the Hydrogen-Air
chemistry from~\citep{Oco04}. The chosen domain is $6$ cm in length
with $h=\expnumber 5{-5}\,$m. We initialize the domain to an ambient
pressure of $1$ bar, temperature of $300$ K, $X_{H_{2}}=0.188$,
$X_{O_{2}}=0.170$, and $X_{N_{2}}=0.642$. A small section on the
right hand side of the domain is initialized to the fully reacted
state from Cantera's homogeneous constant pressure reactor simulation~\citep{cantera}. 

The right hand side boundary is fixed to constant pressure with remaining
variables interpolated from the interior state. The left hand side
boundary is a characteristic wave boundary condition that allows any
pressure waves caused by the initialization to exit the domain. The
initialization contains a temperature and species discontinuity which
gives rise to a pressure oscillation, which is supported by the analysis
in Section~\ref{sec:Discontinuities}. The initial pressure oscillations
leave the system as the reaction front diffuses into the unreacted
region, which eventually creates a stable propagating flame. 

\begin{figure}[H]
\subfloat[Temperature profile, solid line is the Cantera solution on a uniform
$h=\protect\expnumber 1{-5}\,$m grid, dashed grey line is the $\mathrm{DG}\left(p=1\right)$
solution. Results are shifted so that $T=400$ K at $x=0$ for both
$\mathrm{DG}\left(p=1\right)$ and Cantera solutions. \label{fig:flame_T_p1}]{\includegraphics[width=0.45\columnwidth]{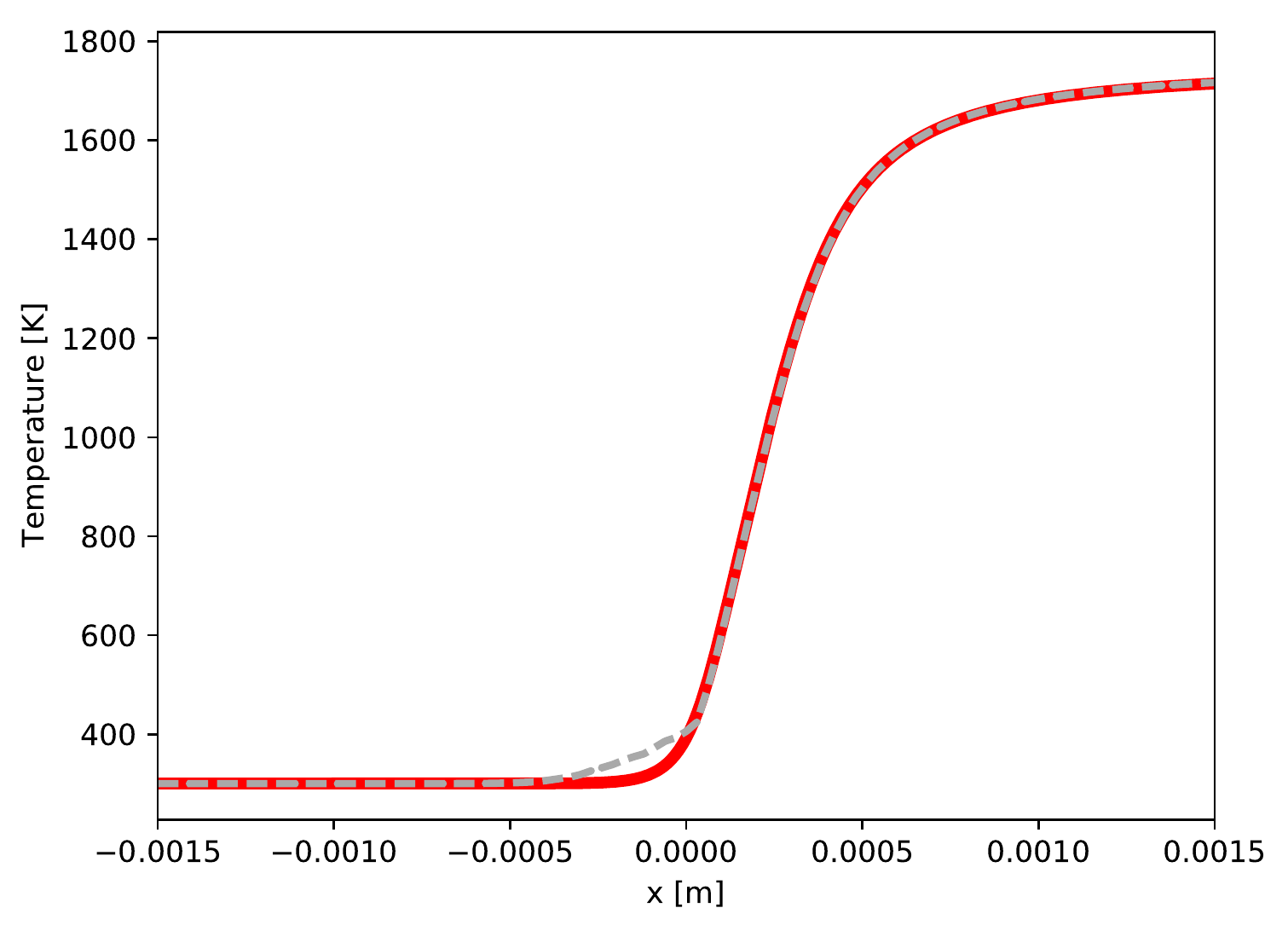}}\hfill{}\subfloat[Minor species profile, solid lines are Cantera solution on a uniform
$h=\protect\expnumber 1{-5}\,$m grid, neighboring dashed grey lines
are the$\mathrm{DG}\left(p=1\right)$ solution. Results are shifted
so that $T=400$ K at $x=0$ for both $\mathrm{DG}\left(p=1\right)$
and Cantera solutions. \label{fig:flame_minor_p1}]{\includegraphics[width=0.45\columnwidth]{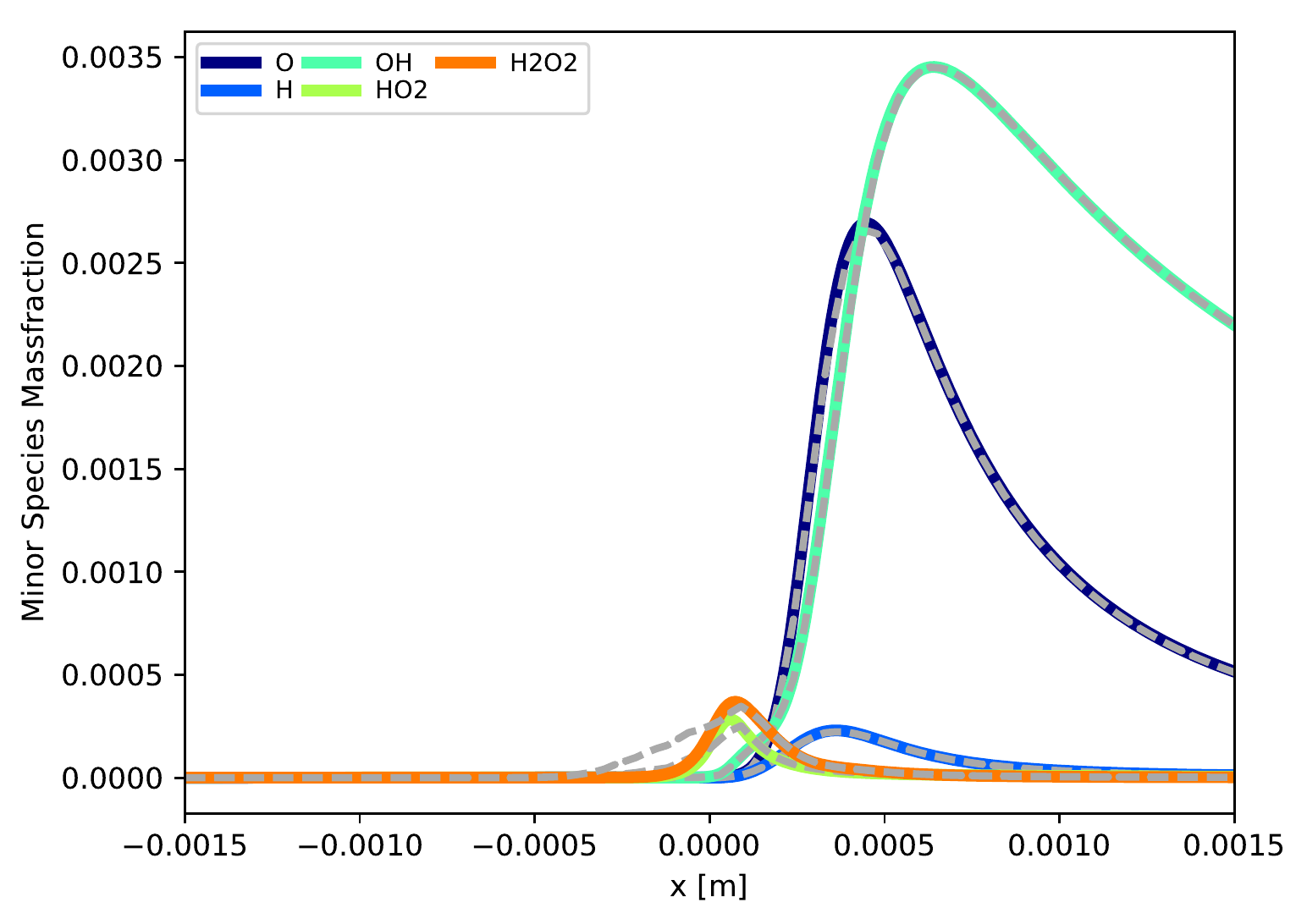}}\hfill{}\subfloat[Major species profile, solid lines are Cantera solution on a uniform
$h=\protect\expnumber 1{-5}\,$m grid, neighboring dashed grey lines
are the $\mathrm{DG}\left(p=1\right)$ solution. Results are shifted
so that $T=400$ K at $x=0$ for both $\mathrm{DG}\left(p=1\right)$
and Cantera solutions.\label{fig:flame_major_p1-2}]{\includegraphics[width=0.45\columnwidth]{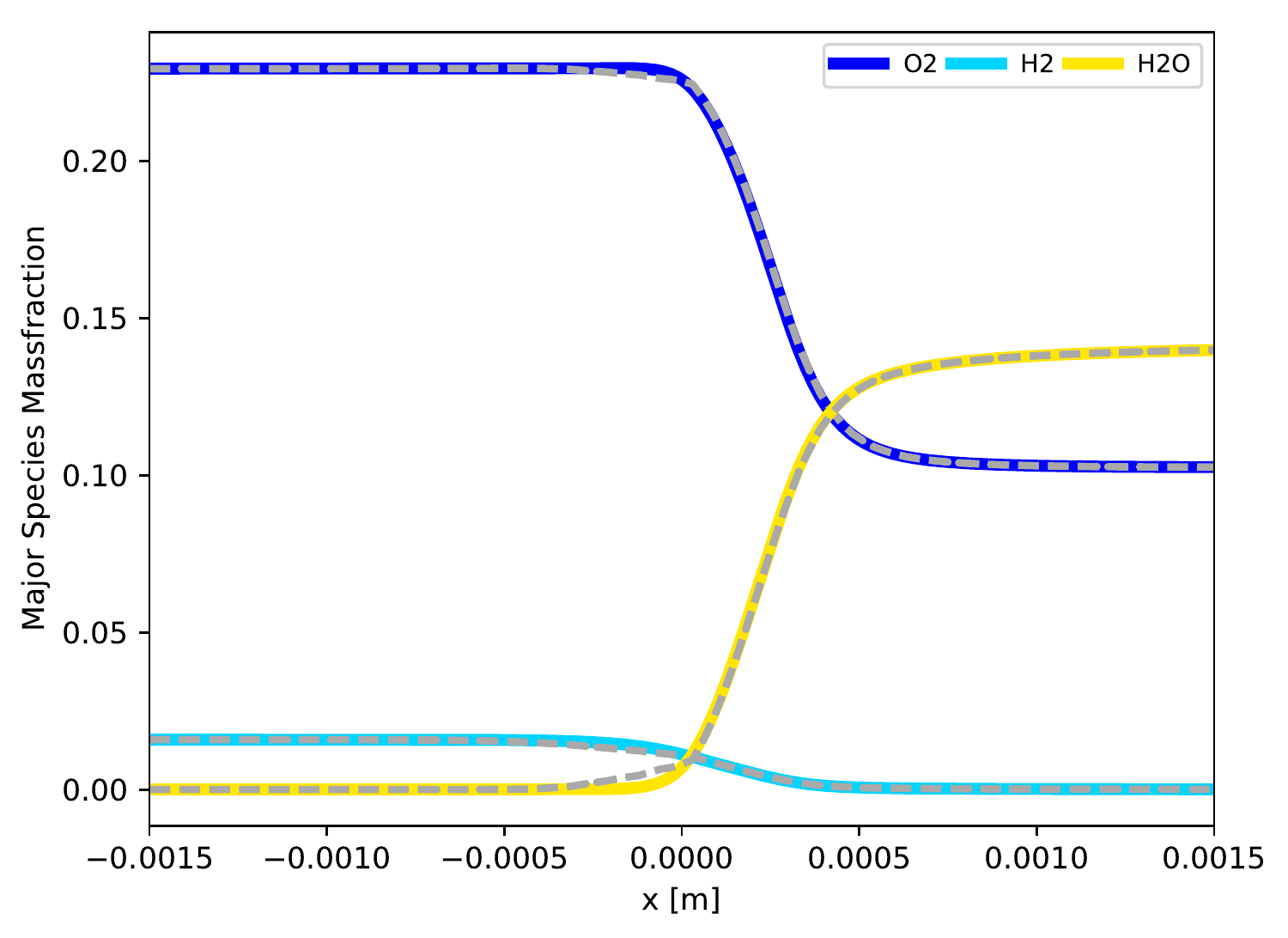}}\hfill{}\subfloat[Pressure profile for the entire simulation domain, solid line is constant
1 bar pressure, dashed grey line is $\mathrm{DG}\left(p=1\right)$
solution. \label{fig:flame_p_p1}]{\includegraphics[width=0.45\columnwidth]{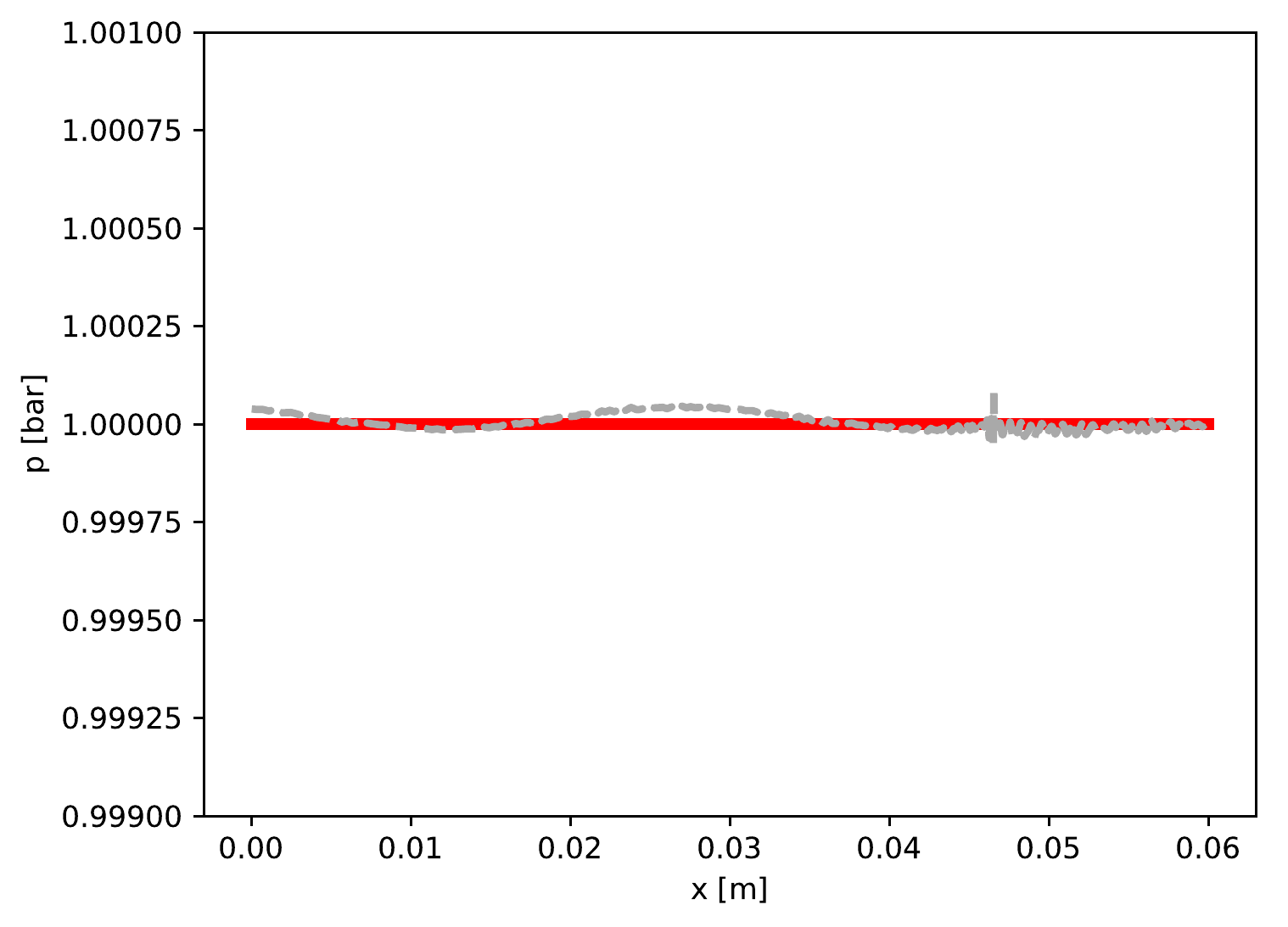}}

\caption{$\mathrm{DG}\left(p=1\right)$ flame solution \label{fig:flame_p1}}
\end{figure}

\begin{figure}[H]
\subfloat[Temperature profile, solid line is the Cantera solution on a uniform
$h=\protect\expnumber 1{-5}\,$m grid, dashed grey line is the $\mathrm{DG}\left(p=2\right)$
solution. Results are shifted so that $T=400$ K at $x=0$ for both
$\mathrm{DG}\left(p=2\right)$ and Cantera solutions.\label{fig:flame_T_p2}]{\includegraphics[width=0.45\columnwidth]{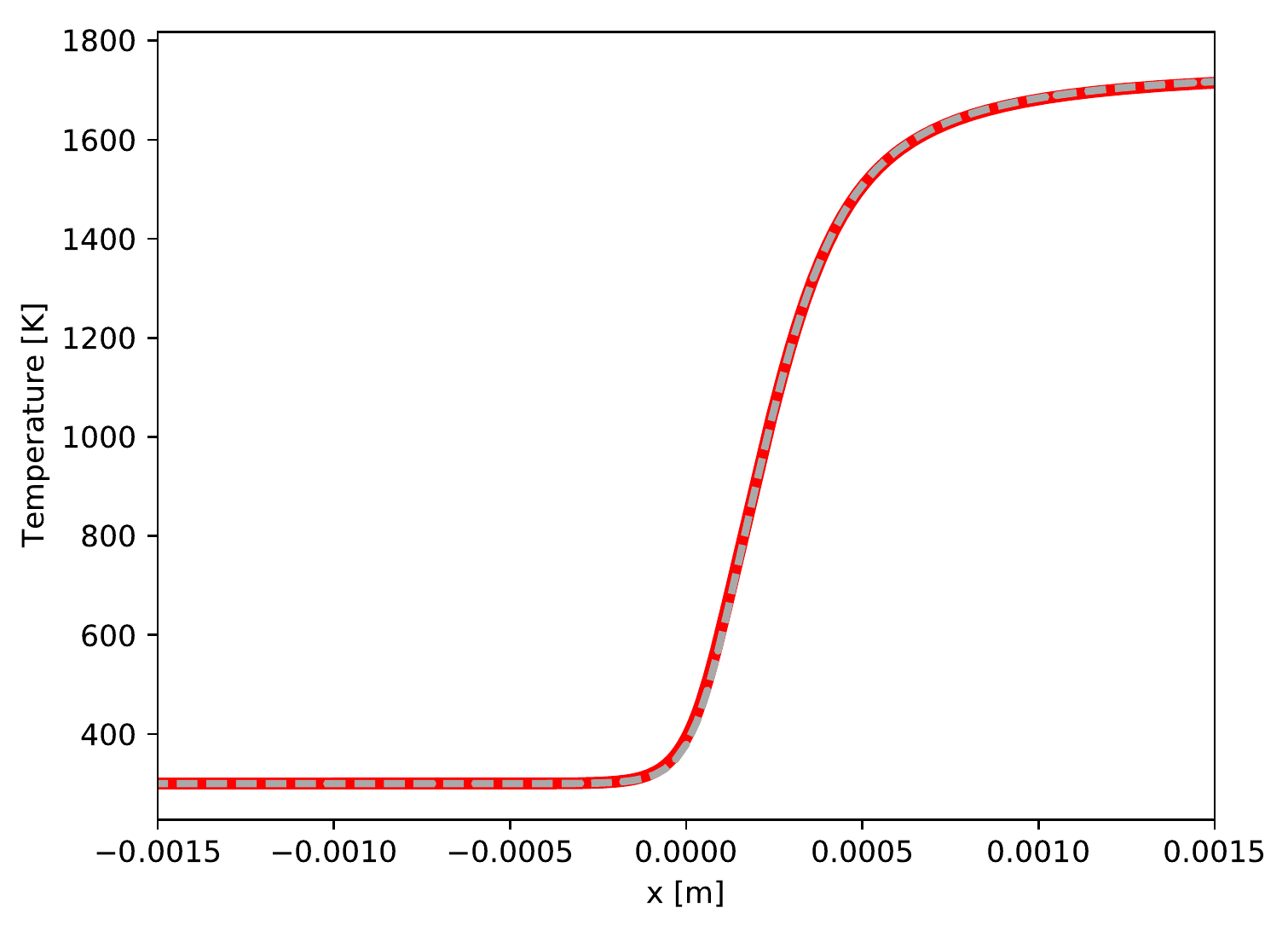}}\hfill{}\subfloat[Minor species profile, solid lines are Cantera solution on a uniform
$h=\protect\expnumber 1{-5}\,$m grid, neighboring dashed grey lines
are the $\mathrm{DG}\left(p=2\right)$ solution. Results are shifted
so that $T=400$ K at $x=0$ for both $\mathrm{DG}\left(p=2\right)$
and Cantera solutions.\label{fig:flame_minor_p2}]{\includegraphics[width=0.45\columnwidth]{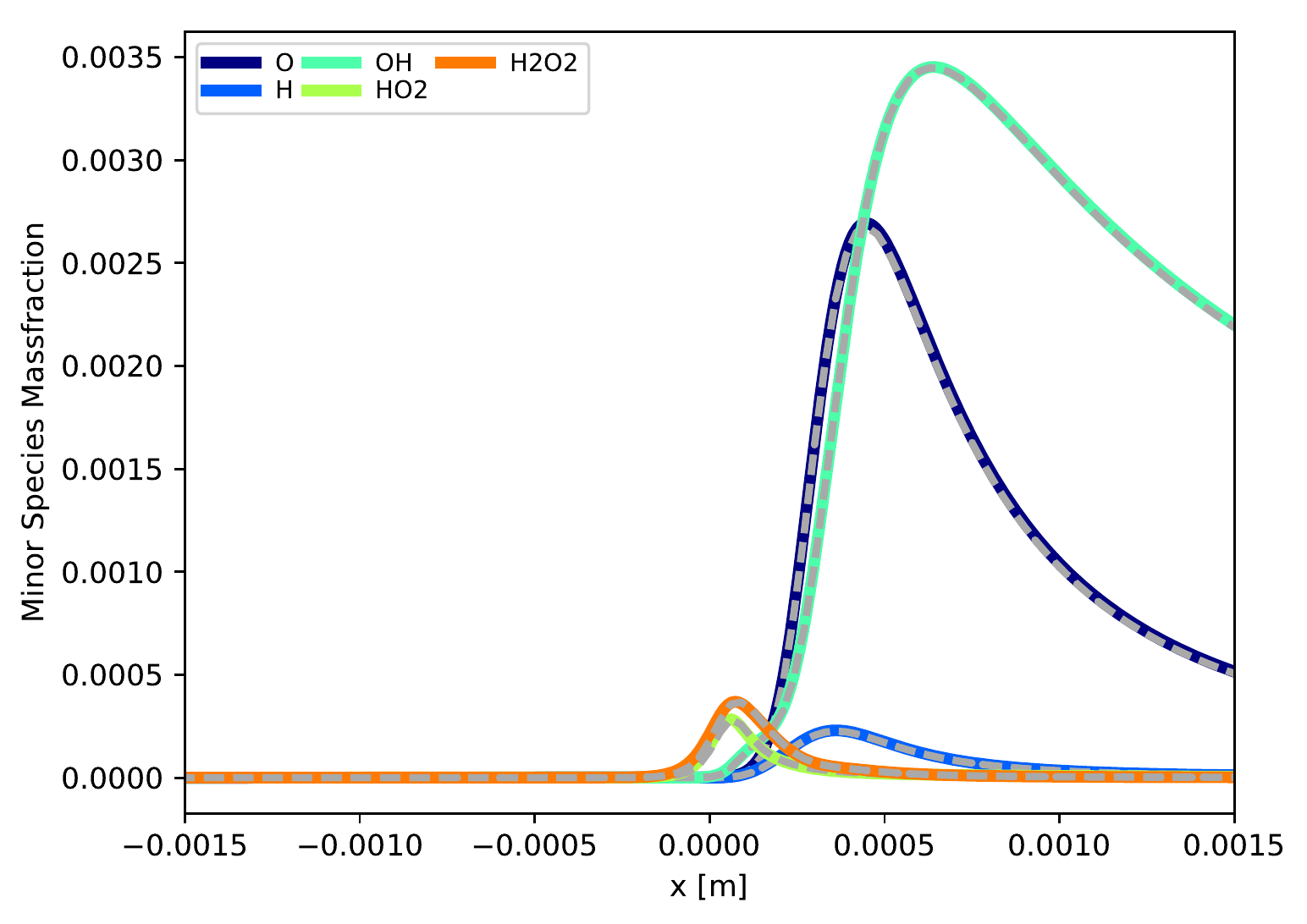}}\hfill{}\subfloat[Major species profile, solid lines are Cantera solution on a uniform
$h=\protect\expnumber 1{-5}\,$m grid, neighboring dashed grey lines
are the $\mathrm{DG}\left(p=2\right)$ solution. Results are shifted
so that $T=400$ K at $x=0$ for both $\mathrm{DG}\left(p=2\right)$
and Cantera solutions.\label{fig:flame_major_p2}]{\includegraphics[width=0.45\columnwidth]{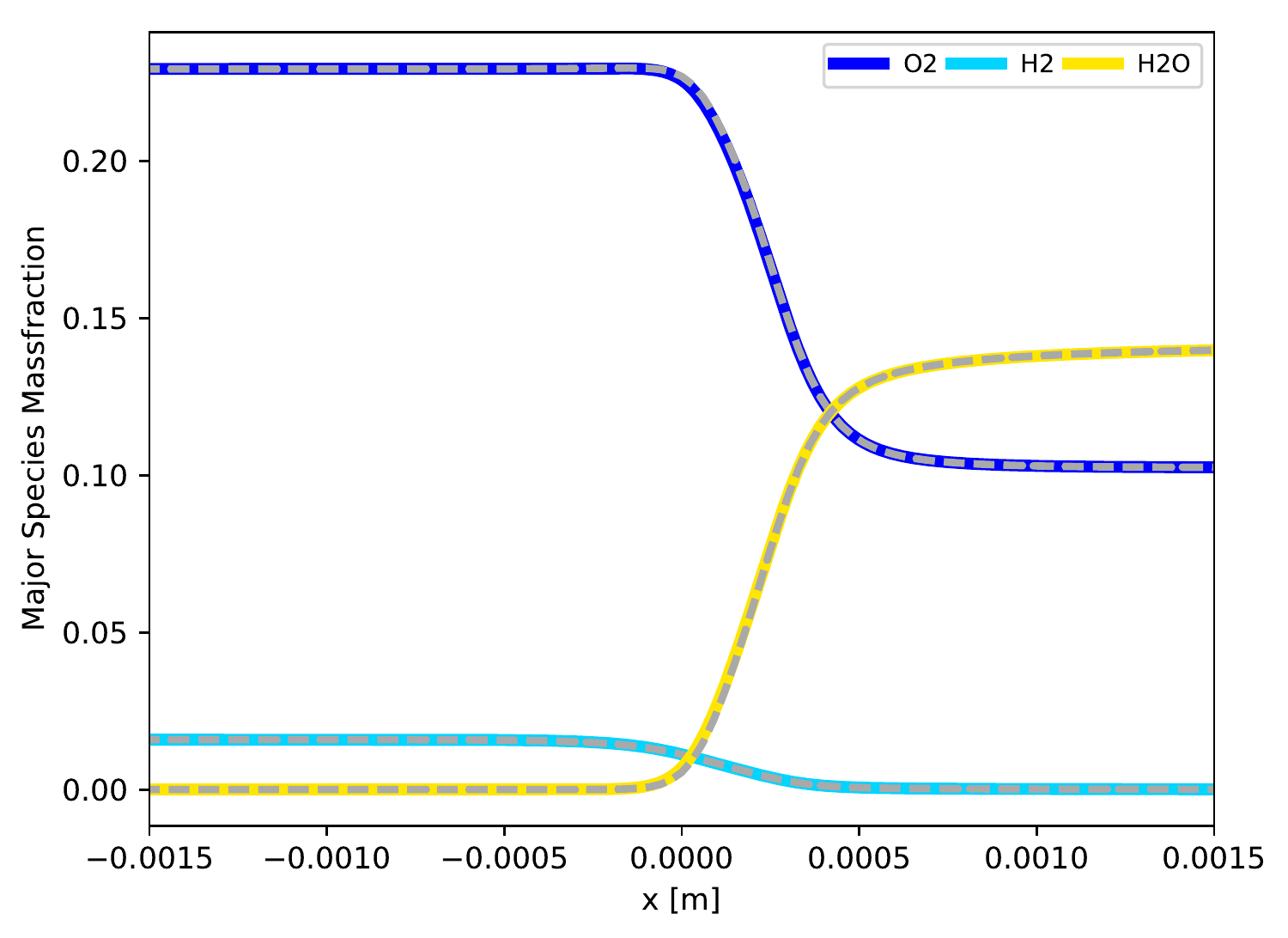}}\hfill{}\subfloat[Pressure profile for the entire simulation domain, solid line is constant
1 bar pressure, dashed grey line is $\mathrm{DG}\left(p=2\right)$
solution.\label{fig:flame_p_p2}]{\includegraphics[width=0.45\columnwidth]{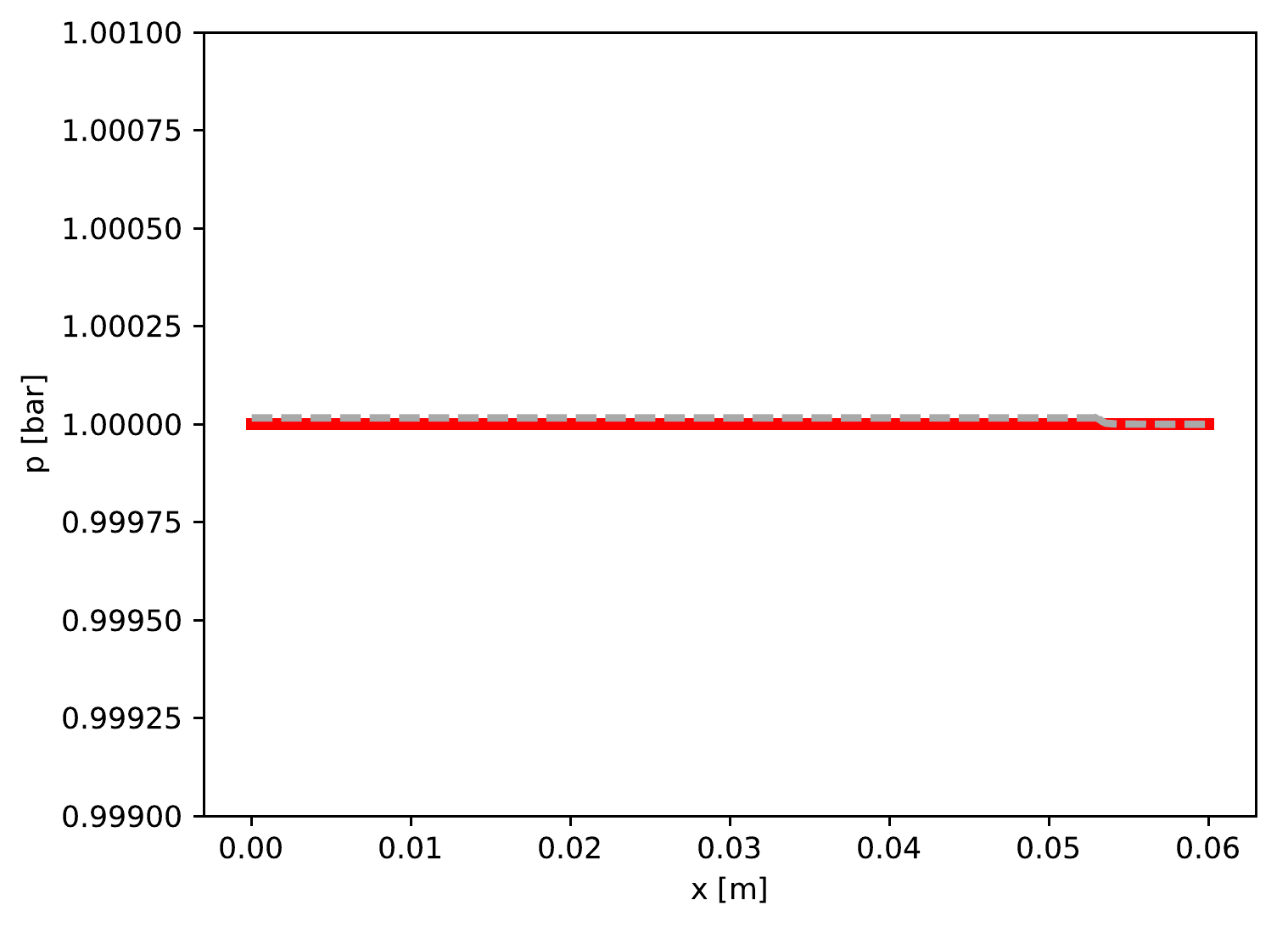}}

\caption{$\mathrm{DG}\left(p=2\right)$ flame solution\label{fig:flame_p2}}
\end{figure}

Figs.~\ref{fig:flame_p1} and \ref{fig:flame_p2} show solutions
for the $\mathrm{DG}\left(p=1\right)$ and $\mathrm{DG}\left(p=2\right)$
solutions, respectively. The temperature and species mass fraction
profiles are compared to the Cantera flame solution with $h=\expnumber 1{-5}\,$m
and the profiles are shifted to have $T=400$ K at $x=0$. The $\mathrm{DG}\left(p=1\right)$
solutions reach the correct reacted state but cannot fully resolve
the flame structures in the $-0.0005<x<0$ region. The $\mathrm{DG}\left(p=2\right)$
solution overcomes these errors and is in good agreement with the
Cantera solution. Despite the under resolved profiles in the $\mathrm{DG}\left(p=1\right)$
solution, both solutions come close to the flame speed calculated
in Cantera. The Cantera flame speed, given as the inflow velocity
for the constant mass flow-rate, is 0.643 m/s. We considered the flame
front in the unsteady $\mathrm{DG}\left(p=1\right)$ and $\mathrm{DG}\left(p=2\right)$
solutions to be the location corresponding to $T=1000$K. We tracked
this location and computed a steady velocity of 0.641 m/s and 0.643
m/s for the $\mathrm{DG}\left(p=1\right)$ and $\mathrm{DG}\left(p=2\right)$
solutions, respectively.

Figs.~\ref{fig:flame_p_p1} and \ref{fig:flame_p_p2} show the pressure
through the entire computational domain for $\mathrm{DG}\left(p=1\right)$
and $\mathrm{DG}\left(p=2\right)$, respectively. For the $\mathrm{DG}\left(p=1\right)$
solution, there are small oscillations, on the order of $0.25$\%
of the ambient pressure. These oscillations are not present for the
$\mathrm{DG}\left(p=2\right)$ solution, where only a slight variation
is seen through the flame front but is constant on both sides of the
flame within $0.1$\% of the desired ambient pressure. The lack of
pressure oscillations in the higher order solution indicate that the
pressure oscillations in the $\mathrm{DG}\left(p=1\right)$ solution
are due to the under-resolved flame and not related to the instabilities
that would require stabilization. 


\section{Conclusion}

We have presented the total energy formulation and the ratio of specific
heats formulation of the reacting Navier-Stokes equations and examined
pressure oscillations for material interfaces with discontinuities
in species and temperature. We derived that if the species internal
energy is nonlinear with respect to temperature then pressure oscillations
are only generated at material interfaces with discontinuities in
both species and temperature. Pressure oscillations are not generated
at a material interface with a species discontinuity if the temperature
is continuous through the interface unless the thermodynamics are
frozen during the temporal integration of the conserved state. The
total energy formulation does not freeze the thermodynamics; instead
a nonlinear relationship for temperature is solved which ensures consistency
between the internal energy of the conserved state and the internal
energy defined by a mixture averaged polynomial expression based on
temperature. The converged temperature can then be use to evaluate
other thermodynamic quantities so that they are consistent with the
conserved state. As such, the total energy formulation can be integrated
with a fully conservative method without generating pressure oscillations
in regions of continuous pressure and temperature.

We presented several test cases to demonstrate the generation of unphysical
pressure oscillations. We demonstrated that the pressure oscillations
from the ratio of specific heats formulation with frozen $\bar{C}_{p}$
did not reach the magnitudes previously reported in literature ~\citep{Abg01,Bil03,Bil11,Hou11,Lv15}.
However, when it was assumed that the mean specific heat at constant
pressure was the same as the NASA polynomial specific heat at constant
pressure, $\bar{C_{p}}=C_{p}$, the oscillations did reach the same
levels as previously reported, indicating that the severity of the
pressure oscillations is dependent on the correct evaluation of thermodynamic
quantities. Regardless, the pressure should be constant across the
material interfaces and when $\bar{C}_{p}$ is frozen the magnitude
of pressure oscillations will grow in time as the solution evolves.
Additionally, the expected numerical pressure oscillations due to
temperature discontinuities at material interfaces were less for the
total energy formulation than for the ratio of specific heats formulation
with frozen $\bar{C}_{p}$.

Previous work showed that advecting continuous species and temperature
profiles caused unstable oscillations~\citep{Bil11,Lv15}. We analyzed
continuous profiles by presenting the thermal bubble test case for
a multicomponent mixture of hydrogen and oxygen from ~\citep{Lv15}.
The total energy formulation solution for this test case did not generate
pressure oscillations. The solutions reached the magnitudes of pressure
oscillations found in the previous work only when the ratio of specific
heats with frozen $C_{p}$ was used. When the ratio of specific heats
with frozen $\bar{C}_{p}$ was used no pressure oscillations were
generated due to the lack of numerical mixing for an advected continuous
profile of species and temperature.

Finally, we presented the results for a fully reacting Navier-Stokes
simulation of a continuous one-dimensional flame. The solution, computed
using a fully conservative $\mathrm{DG}$ method without additional
stabilization, compared well with the Cantera solution. Flame speeds
for both the $\mathrm{DG}(p=1)$ and $\mathrm{DG}(p=2)$ solutions
were consistent with the Cantera flame speed. Discrepancies between
the Cantera solution and the $\mathrm{DG}(p=1)$ solution are not
observed in the $\mathrm{DG}(p=2)$ solution, indicating the $\mathrm{DG}(p=2)$
solution better resolved the flame.

The total energy formulation does not require non-conservative methods
or additional stabilization in smooth regions of the flow. This is
an attractive feature since both non-conservative methods and artificial
stabilization have associated inherent costs. Non-conservative methods
often involve additional cell and face loops that increase the computational
complexity. Additionally, artificial stabilization can prevent high
order methods from achieving the formal order accuracy associated
with the polynomial space of the approximate solution. We also demonstrated
that the total energy formulation is equivalent to the ratio of specific
heats formulation in non-reacting regions of the flow where thermodynamic
quantities are not changing, which is a suitable alternative and may
be convenient for methods that require the internal energy to be defined
according to the ratio of specific heats, e.g., characteristic boundary
conditions.

\bibliographystyle{elsarticle-num}
\bibliography{citations}

\end{document}